\documentclass[twocolumn,twoside]{IEEEtran}

\usepackage{amsmath,amstext,amssymb,epsf}

 \newtheorem{lemma}{Lemma}[section]
 \newtheorem{theorem}[lemma]{Theorem}
 \newtheorem{proposition}[lemma]{Proposition}
 
 \newtheorem{claim}[lemma]{Claim}
 \newtheorem{corollary}[lemma]{Corollary}

 \newtheorem{definition}[lemma]{Definition}
\newtheorem{rem}[lemma]{Remark}
\newenvironment{remark}{\begin{rem}}{\hspace*{\fill}$\diamondsuit$\end{rem}}
 \newtheorem{ex}[lemma]{Example}
\newenvironment{example}{\begin{ex}}{\hspace*{\fill}$\diamondsuit$\end{ex}}

\numberwithin{equation}{section} 

\newcommand{\eps}{\varepsilon}

\newcommand{\pair}[1]{\langle #1\rangle}

\newcommand{\N}{\mathbb{N}}

\newcommand{\len}[2]{l_{#1}(#2)}
\newcommand{\close}[2]{#1={\mathcal E}(#2)}
\newcommand{\dclose}[2]{#1 \cong #2}

\newcommand{\Loss}{{\rm Loss}}

\begin{document}
\title{Kolmogorov's Structure Functions and \\ Model Selection}
\author{Nikolai Vereshchagin
and Paul Vit\'anyi\thanks{Manuscript received xxx, 2002;
revised yyy 2003. The material of this paper was presented in part in
{\em Proc. 47th IEEE Symp. Found. Comput. Sci.}, 2002, 751-760. 
Part of this work was done during N.K. Vershchagin's stay at CWI.
His work was supported in part by Grant 
01-01-01028 from Russian Federation
Basic Research Fund, by the Netherlands Organization for
Scientific Research (NWO) Grant EW 612.052.004, and by CWI.
The work of P.M.B. Vit\'anyi was supported in part
by the EU fifth framework project QAIP, IST--1999--11234, 
the NoE QUIPROCONE IST--1999--29064,
the ESF QiT Programmme, and the EU Fourth Framework BRA
NeuroCOLT II Working Group
EP 27150, and the EU NoE PASCAL. 
N.K. Vereshchagin is with the
Department of Mathematical Logic and Theory of Algorithms,
Faculty of Mechanics and Mathematics, Moscow State University,
Leninskie Gory, Moscow, Russia 119992. Email:
{\tt ver@mech.math.msu.su}.
P.M.B. Vit\'anyi is with 
CWI, Kruislaan 413,
1098 SJ Amsterdam, The Netherlands. 
Email: {\tt Paul.Vit\'anyi@cwi.nl}.}
}
\markboth{IEEE Transactions on Information Theory, VOL. XX, NO Y, MONTH 2003}{Vereshchagin and Vit\'anyi: Kolmogorov's Structure Functions and Model Selection}

\maketitle

\begin{abstract}
In 1974 Kolmogorov proposed a non-probabilistic approach to statistics
and model selection. Let data be finite binary strings and models be
finite sets of binary strings. Consider model classes
consisting of models of given maximal (Kolmogorov) complexity.
The ``structure function'' of the given data expresses the
relation between the complexity level constraint on a model class
and the least log-cardinality of a model in the class containing the data.  
We show that the structure function determines all stochastic properties
of the data: for every constrained model class it determines the individual
best-fitting model in the class irrespective of whether the ``true'' model
is in the model class considered or not.
In this setting, this happens {\em with certainty}, 
rather than with high probability as is in the classical case.
We precisely quantify the goodness-of-fit of an individual
model with respect to individual data.
We show that---within the obvious constraints---every graph 
is realized by the structure function of some data.
We determine the  (un)computability properties of the
various functions contemplated  and of the ``algorithmic 
minimal sufficient statistic.''


\end{abstract}

\section{Introduction}
\label{sect.intro}
As perhaps the last mathematical innovation of an extraordinary
scientific career, 
A.N. Kolmogorov \cite{Ko74a,Ko74b}
proposed to found statistical theory on finite combinatorial
principles independent of probabilistic assumptions.
Technically, the new statistics is expressed in terms of Kolmogorov
complexity, \cite{Ko65}, the information in an individual object.
The relation
between the individual data and its explanation (model) is expressed
by Kolmogorov's structure function.
This function,
its variations and its relation to model selection,
have obtained some notoriety 
\cite{Sh83,Co85,Vy87,CT91,LiVi97,Sh99,Vy99,GTV01,Le02,Ga02,Co02},
but it has not before been
comprehensively analyzed
and understood.
It has often been questioned
why Kolmogorov chose to focus on the
the mysterious function $h_x$ below,
rather than on
the more evident $\beta_x$ variant below. The only written record
by Kolmogorov himself is the following abstract \cite{Ko74b}
(translated from the original Russian by L.A. Levin):

\label{sect.trans}
``To each constructive object corresponds a function $\Phi_x(k)$ of a
 natural number $k$---the log of minimal cardinality of $x$-containing
 sets that allow definitions of complexity at most $k$.
 If the element $x$ itself allows a simple definition,
 then the function $\Phi$ drops to $1$ 
 even for small $k$.
 Lacking such definition, the element is ``random'' in a negative sense.
 But it is positively ``probabilistically random'' only when function
 $\Phi$ having taken the value $\Phi_0$ at a relatively small
 $k=k_0$, then changes approximately as $\Phi(k)=\Phi_0-(k-k_0)$.''

These pregnant lines will become clear on reading this paper,
where we use ``$h_x$'' for the structure function ``$\Phi_x$.''
Our main result establishes the
importance of the  structure function:
For every data item, and every complexity level, 
minimizing a two-part code, 
one part model description and one part data-to-model code
(essentially a constrained two-part MDL estimator \cite{Ri83}),
over the class of models of at most the given complexity,
{\em with certainty}
(and not only with high probability) selects models that
in a rigorous sense are the best explanations  among the contemplated models.
The same holds for minimizing the one-part code consisting of
just the data-to-model code
(essentially a constrained maximum likelihood estimator).
The explanatory value of an individual model for particular data,
its goodness of fit, is quantified by
by the randomness deficiency
\eqref{eq:randomness-deficiency} expressed in terms of
Kolmogorov complexity: minimal randomness deficiency implies
that the data is maximally ``random'' or ``typical'' for the model. 
It turns out that the minimal randomness deficiency of 
the data in
a complexity-constrained model class
cannot be computationally monotonically approximated (in the sense
of Definition~\ref{def.semi}) up 
to any significant precision.
Thus, while
we can monotonically approximate (in the precise sense of
Section~\ref{sect.real})
the minimal length two-part code, or the one-part code,
and thus monotonically approximate {\em implicitly} 
the best fitting model, 
we cannot monotonically approximate the number expressing the goodness
of this fit. 
But this should be sufficient: we want the best model
rather than a number that measures its goodness.  

\subsection{Randomness in the Real World} 
Classical statistics investigates real-world phenomena
using probabilistic methods.
There is the problem of what probability means,
whether it is subjective, objective, or exists at all.
P.S. Laplace conceived of the probability of a physical
event as expressing lack of knowledge concerning its true
deterministic causes \cite{La19}. 
A. Einstein rejected physical random variables as well 
``I do not believe
that the good Lord plays dice.'' 
But even if true physical random variables do exist,
can we assume that a particular phenomenon we want to explain is
probabilistic? Supposing that to be the case as well, we then
use a probabilistic
statistical method to select models. In this situation
the proven ``goodness'' of such a method
is so only in a probabilistic sense.
But for current applications, 
the total probability concentrated
on potentially realizable data may be negligible,
for example, in complex video and sound
data. In such a case,
a model selection process that is successful with high probability 
may
nonetheless fail on the actually realized data.
Avoiding these difficulties, Kolmogorov's proposal
strives
for the firmer and less contentious ground of finite combinatorics
and effective computation.

\subsection{Statistics and Modeling}
Intuitively, a central task of statistics is to identify the true source
that produced the data at hand. But suppose the true source
is 100,000 fair coin flips and our data is the outcome $00 \ldots 0$. 
A method that
identifies flipping a fair coin as the cause of this outcome is 
surely a bad method, even though the source of the data 
it came up with happens
to be the true cause. 
Thus, for a good statistical method to work well we assume that
the data are ``typical'' for the source that produced the data, 
so that the source ``fits'' the data.
The situation is more subtle for data like
$0101 \ldots 01$. Here the outcome of the source has an equal frequency
of 0s and 1s, just like we would expect from a fair coin.
But again, it is virtually impossible that such data are produced by
a fair coin flip, or indeed, independent flips of a coin of any 
particular bias. 
In real-world phenomena
we cannot be sure that the true source of the data is
in the class of sources considered, or, worse, we are virtually certain 
that the true source is not in that class.
Therefore, the real question is not to find the true cause
of the data, but to model the data as well as possible. 
In recognition of this, we often talk about ``models'' instead of ``sources,'' 
and the contemplated ``set of sources'' is called the 
contemplated ``model class.''
In traditional statistics ``typicality'' and ``fitness' are probabilistic
notions tied to sets of data and models of large measure. In the
Kolmogorov complexity setting we can express and quantify ``typicality''
of individual data with respect to a single model, and express and quantify 
the ``fitness'' of an individual model
for the given data.

\section{Preliminaries}

Let $x,y,z \in {\cal N}$, where
${\cal N}$ denotes the natural
numbers and we identify
${\cal N}$ and $\{0,1\}^*$ according to the
correspondence
\[(0, \epsilon ), (1,0), (2,1), (3,00), (4,01), \ldots \]
Here $\epsilon$ denotes the {\em empty word}.
The {\em length} $|x|$ of $x$ is the number of bits
in the binary string $x$, not to be confused with the {\em cardinality}
$|S|$ of a finite set $S$. For example,
$|010|=3$ and $|\epsilon|=0$, while $|\{0,1\}^n|=2^n$ and
$|\emptyset|=0$.
The emphasis is on binary sequences only for convenience;
observations in any alphabet can be so encoded in a way
that is `theory neutral'.
Below we will use the natural numbers and the binary strings
interchangeably. 

\subsection{Self-delimiting Code}
A binary string $y$
is a {\em proper prefix} of a binary string $x$
if we can write $x=yz$ for $z \neq \epsilon$.
 A set $\{x,y, \ldots \} \subseteq \{0,1\}^*$
is {\em prefix-free} if for any pair of distinct
elements in the set neither is a proper prefix of the other.
A prefix-free set is also called a {\em prefix code} and its
elements are called {\em code words}. 
An example of a
prefix code, that is useful later, 
encodes the source word $x=x_1 x_2 \ldots x_n$ by the code word
\[ \overline{x} = 1^n 0 x .\]
This prefix-free code
is called {\em self-delimiting}, because there is fixed computer program
associated with this code that can determine where the
code word $\bar x$ ends by reading it from left to right without
backing up. This way a composite code message can be parsed
in its constituent code words in one pass, by the computer program.
(This desirable property holds for every prefix-free
encoding of a finite set of source words, but not for every
prefix-free encoding of an infinite set of source words. For a single
finite computer program to be able to parse a code message the encoding needs
to have a certain uniformity property like the $\overline{x}$ code.)
Since we use the natural numbers and the binary strings interchangeably,
$|\bar x|$ where $x$ is ostensibly an integer, means the length
in bits of the self-delimiting code of the binary string with index $x$.
On the other hand, $\overline{|x|}$ where $x$ is ostensibly a binary
string, means the self-delimiting code of the binary string
with index the length $|x|$ of $x$.   
Using this code we define
the standard self-delimiting code for $x$ to be
$x'=\overline{|x|}x$. It is easy to check that
$|\overline{x} | = 2 n+1$ and $|x'|=n+2 \log n +1$.
Let $\langle \cdot \rangle$ denote a standard invertible
effective one-one encoding from ${\cal N} \times {\cal N}$
to a subset of ${\cal N}$.
For example, we can set $\langle x,y \rangle = x'y$
or $\langle x,y \rangle = \bar xy$.
We can iterate this process to define
$\langle x , \langle y,z \rangle \rangle$,
and so on.

\subsection{Kolmogorov Complexity}
For precise definitions, notation, and results see the text \cite{LiVi97}.
Informally, the Kolmogorov complexity, or algorithmic entropy, $K(x)$ of a
string $x$ is the length (number of bits) of a shortest binary
program (string) to compute
$x$ on a fixed reference universal computer
(such as a particular universal Turing machine).
Intuitively, $K(x)$ represents the minimal amount of information
required to generate $x$ by any effective process.
The conditional Kolmogorov complexity $K(x | y)$ of $x$ relative to
$y$ is defined similarly as the length of a shortest program
to compute $x$, if $y$ is furnished as an auxiliary input to the
computation.
For technical reasons we use a variant of complexity,
so-called prefix complexity, which is associated with Turing machines
for which the set of programs resulting in a halting computation
is prefix free. 
We realize prefix complexity by considering a special type of Turing
machine with a one-way input tape, a separate work tape,
and a one-way output tape. Such Turing
machines are called {\em prefix} Turing machines. If a machine $T$ halts
with output $x$
after having scanned all of $p$ on the input tape, 
but not further, then $T(p)=x$ and
we call $p$ a {\em program} for $T$.
It is easy to see that 
$\{p : T(p)=x, x \in \{0,1\}^*\}$ is a {\em prefix code}.
Let $T_1 ,T_2 , \ldots$ be a standard enumeration
of all prefix Turing machines with a binary input tape,
for example the lexicographical length-increasing ordered syntactic
prefix Turing machine descriptions, \cite{LiVi97},
and let $\phi_1 , \phi_2 , \ldots$
be the enumeration of corresponding functions
that are computed by the respective Turing machines
($T_i$ computes $\phi_i$).
These functions are the {\em partial recursive} functions
or {\em computable} functions (of effectively prefix-free encoded
arguments). The Kolmogorov complexity
of $x$ is the length of the shortest binary program
from which $x$ is computed. 
\begin{definition}\label{def.KolmK}
The {\em prefix Kolmogorov complexity} of $x$ is
                  \begin{equation}\label{eq.KC}
K(x) = \min_{p,i}  \{|\bar i| +  |p|:
T_i (p)=x \} , 
                  \end{equation}
where the minimum is taken over $p \in \{0,1\}^*$ and  $i 
\in \{1,2, \ldots \}$. 
For the development of the theory we
actually require  
the Turing machines to use {\em auxiliary} (also
called {\em conditional})
information, by equipping the machine with a special
read-only auxiliary tape containing this information at the outset.
Then, the {\em conditional version} $K(x \mid y)$ of the prefix
Kolmogorov complexity of $x$  
given $y$ (as
auxiliary information) is
is defined similarly as before,
and the unconditional version is set to  $K(x)=K(x  \mid \epsilon)$.
\end{definition}

One of the main achievements of the theory of computation
is that the enumeration $T_1,T_2, \ldots$ contains 
a machine, say $U=T_u$, that is computationally universal in that it can
simulate the computation of every machine in the enumeration when
provided with its index:
    $U(\langle y, \bar{i}p)  = T_i (\langle y,p\rangle)$
    for all $i,p,y$.
    We fix one such machine and designate it as the {\em reference universal
    prefix Turing machine}.
    Using this universal machine it is easy to show
    $K (x \mid y) = \min_q \{|q|: U(\langle y,q\rangle)=x \}$.

A prominent property of the prefix-freeness of $K(x)$ is 
that we can interpret $2^{-K(x)}$
as a probability distribution since $K(x)$ is the length of
a shortest prefix-free program for $x$. By the fundamental
Kraft's inequality, see for example \cite{CT91,LiVi97}, we know that
if $l_1 , l_2 , \ldots$ are the code-word lengths of a  prefix code,
then $\sum_x 2^{-l_x} \leq 1$. Hence, 
\begin{equation}\label{eq.udconv}
\sum_x 2^{-K(x)} \leq 1.
\end{equation}
This leads to the notion
of universal distribution---a rigorous form of Occam's razor---which
implicitly plays an important part in the present exposition.
  The functions $K( \cdot)$ and $K( \cdot \mid  \cdot)$,
though defined in terms of a
particular machine model, are machine-independent up to an additive
constant
 and acquire an asymptotically universal and absolute character
through Church's thesis, from the ability of universal machines to
simulate one another and execute any effective process.
  The Kolmogorov complexity of an individual object was introduced by
Kolmogorov \cite{Ko65} as an absolute
and objective quantification of the amount of information in it.
The information theory of Shannon \cite{Sh48}, on the other hand, 
deals with {\em average} information {\em to communicate}
objects produced by a {\em random source}.
  Since the former theory is much more precise, it is surprising that
analogs of theorems in information theory hold for
Kolmogorov complexity, be it in somewhat weaker form.
An example is the remarkable {\em symmetry of information} property
used later.
Let $x^*$  denote the shortest prefix-free program $x^*$ 
for a finite string $x$,
or, if there are more than one of these, then $x^*$ is the first
one halting in a fixed standard enumeration of all halting programs.
Then, by definition, $K(x)=|x^*|$.
Denote $K(x,y)=K(\langle x,y \rangle)$. Then,
\begin{align}\label{eq.soi}
K(x,y) & = K(x)+K(y \mid x^*) + O(1) \\
& = K(y)+K(x \mid y^*)+O(1) .
\nonumber
\end{align}
\begin{remark}\label{rem.xKx}
\rm
The information contained in $x^*$ in the conditional above
is the same as the information in the pair $(x,K(x))$, up to an additive 
constant, since there are recursive functions $f$ and $g$ such that
for all $x$ we have $f(x^*)=(x,K(x))$ and $g(x,K(x))=x^*$.
On input $x^*$,  the function $f$ computes $x=U(x^*)$ and $K(x)=|x^*|$;
and on input $x,K(x)$ the function $g$ runs all programs of length $K(x)$
simultaneously, round-robin fashion, until the first program computing
$x$ halts---this is by definition $x^*$.
\end{remark}

\subsection{Precision}
It is customary in this area to use ``additive constant $c$'' or
equivalently ``additive $O(1)$ term'' to mean a constant,
accounting for the length of a fixed binary program,  
independent from every variable or parameter in the expression
in which it occurs. In this paper we use the prefix complexity 
variant of Kolmogorov complexity for convenience.
Actually some results, especially Theorem~\ref{theo.th6}, are easier to prove
for plain complexity.  Most results presented here are precise up to
an additive term that is logarithmic in 
the length of the binary string
concerned, 
which means that they
are valid for plain complexity as well---prefix 
complexity of a string exceeds
the plain complexity of that string by at most an 
additive term that is logarithmic in the length of that string. 
Thus, our use of
prefix complexity is important for ``fine details'' only.

\subsection{Meaningful Information}
The information contained in an individual
finite object (like a finite binary string) is measured
by its Kolmogorov complexity---the length of the shortest binary program
that computes the object. Such a shortest program contains no redundancy:
every bit is information; but is it meaningful information?
If we flip a fair coin to obtain a finite binary string, then with overwhelming
probability that string constitutes its own shortest program. However,
also with overwhelming probability all the bits in the string are meaningless
information, random noise. On the other hand, let an object
$x$ be a sequence of observations of heavenly bodies. Then $x$
can be described by the binary
string $pd$, where $p$ is the description of
the laws of gravity, and $d$ the observational
parameter setting:
we can divide the information in $x$ into
meaningful information $p$ and accidental information $d$.
The main task for statistical inference and learning theory is to
distil the meaningful information present in the data. The question
arises whether it is possible to separate meaningful
information from accidental information, and if so, how.
The essence of the solution to this problem is revealed when we
rewrite (\ref{eq.KC})
as follows:
             \begin{align}\label{eq.kcmdl}
K(x) & =
\min_{p,i} \{|\bar i| + |p|:T_i(p) =x\}
\\&  =
\nonumber
\min_{p,i} \{2|i|+|p|+1:T_i(p) =x\}
\\&  \leq
\nonumber
\min_{q} \{|q|: U(\langle \epsilon, q \rangle) =x\}+2|u|+1
\\&  \leq
\nonumber
\min_{r,j} \{K(j)+|r|: U(\langle \epsilon, j^* \alpha r \rangle)
=T_j(r) =x\}
\\&  
\nonumber
\hspace{2.5in} +2|u|+1
\\&  \leq
K(x)+O(1).
\nonumber
\end{align}
Here the minima are taken over
$p,q,r \in \{0,1\}^*$ and $i,j \in \{1,2, \ldots\}$.
The last equalities are obtained by
using the universality of the fixed reference universal prefix Turing machine
$U=T_u$ with $|u| = O(1)$.
The string $j^*$ is a shortest self-delimiting 
program of $K(j)$ bits from which $U$ can compute $j$,
and subsequent execution of the next self-delimiting fixed program
$\alpha$ will compute $\bar j$ from $j$.  Altogether,
this has the effect that $U(\langle \epsilon, j^* \alpha r \rangle)
=T_j(r)$.
This expression emphasizes the two-part code nature of Kolmogorov complexity.
In the example
$$x = 10101010101010101010101010$$
we can encode $x$ by a small Turing machine printing a specified
number of copies of the pattern ``01'' which computes
$x$ from the program ``13.''
This way, $K(x)$ is viewed  as the shortest length of
a two-part code for $x$, one part describing a Turing machine,
or {\em model}, for the {\em regular} aspects of $x$,
and the second part describing
the {\em irregular} aspects of $x$ in the form
of a program to be interpreted by $T$.
The regular, or ``valuable,'' information in $x$ is constituted
by the bits in the ``model'' while the random or ``useless''
information of $x$ constitutes the remainder.

\subsection{Data and Model}
To simplify matters,
and because all discrete data
can be binary coded, we consider only 
finite binary data strings $x$.
Our model class consists
of Turing machines $T$ that enumerate
a finite set, say $S$, such that on input $i \leq |S|$ we have $T(i)=x$
with $x$ the $i$th element of $T$'s enumeration of $S$, and
$T(i)$ is a special {\em undefined} value if $i>|S|$.
The ``best fitting'' model for $x$ is
a Turing machine $T$ that
reaches the minimum description length in (\ref{eq.kcmdl}).
Such a machine $T$ embodies the amount of useful information
contained in $x$, and we have divided 
a shortest program $x^*$ for $x$ into parts $x^*=T^*i$ such
$T^*$ is a shortest self-delimiting program for $T$.
Now suppose
we consider only {\em low} complexity finite-set models,
and under these constraints the shortest two-part description 
happens to be longer than the shortest one-part description.
Does the model minimizing the two-part description
 still capture all (or as much as possible)
meaningful information? Such considerations require
study of the relation between the complexity limit on the contemplated model  
classes, the shortest two-part code length, and the amount
of meaningful information captured.

\subsection{Kolmogorov's Structure Functions}
\begin{figure}
\begin{center}
\epsfxsize=8cm
\epsfxsize=8cm \epsfbox{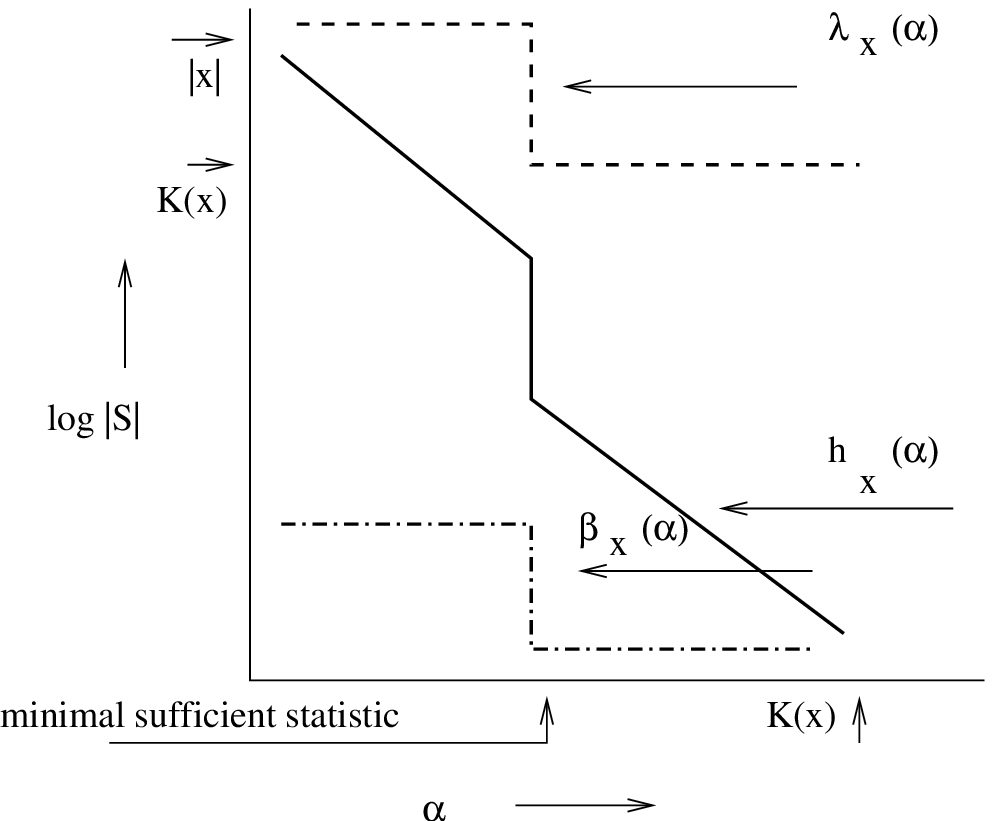}
\end{center}
\caption{Structure functions $h_x(\alpha), \beta_x(\alpha), \lambda_x(\alpha)$,
and minimal sufficient statistic.}
\label{figure.estimator}
\end{figure}

We will prove that there is a close relation between
functions describing
three, a priori seemingly unrelated, aspects of modeling individual
data by models of prescribed complexity: 
optimal fit, 
minimal remaining randomness,
and length of shortest two-part code,
respectively (Figure~\ref{figure.estimator}). We first need a definition.
Denote the {\em complexity
of the finite set} $S$ by
$K(S)$---the length (number of bits) of the
shortest binary program $p$ from which the reference universal
prefix machine $U$
computes a listing of the elements of $S$ and then
halts. 
That is, if $S=\{x_1 , \ldots , x_{n} \}$, then
$U(p)= \langle x_1,\langle x_2, \ldots, \langle x_{n-1},x_n\rangle \ldots\rangle \rangle $.
The shortest program $p$,
or, if there is more than one such shortest program, then
the first one that halts in a standard dovetailed running of all programs,
is denoted by $S^*$. 
The {\em conditional complexity} $K(x \mid S)$ of $x$ given $S$
is the length (number of bits) in the
shortest binary program $p$ from which the reference universal
prefix machine $U$
computes $x$ from input $S$ given literally. 
In the sequel we also use $K(x \mid S^*)$,
defined as the length of the shortest program that
computes $x$ from input $S^*$. Just like in Remark~\ref{rem.xKx},
the input $S^*$ has more
information, namely all information in the pair $(S,K(S))$, 
than just the literal list $S$. 
Furthermore, $K(S \mid x)$ is defined as the 
length of the shortest program that computes $S$ from input $x$,
and similarly we can define $K(S^* \mid x),K(S \mid x^*)$.
For every finite set $S  \subseteq \{0,1\}^*$ containing
$x$ we have 
        \begin{equation}\label{eq57}
K(x  \mid  S)\le\log|S|+O(1).
        \end{equation}
Indeed, consider the selfdelimiting code of $x$
consisting of its $\lceil\log|S|\rceil$ bit long index
of $x$ in the lexicographical ordering of $S$.
This code is called
\emph{data-to-model code}.
Its length quantifies the maximal ``typicality,'' or ``randomness,''
data (possibly different from $x$) can have with respect to this model.
The lack of typicality
of $x$ with respect to $S$
is measured by the amount by which $K(x \mid S)$
falls short of the length of the data-to-model code.
The {\em randomness deficiency} of $x$ in $S$ is defined by
      \begin{equation}\label{eq:randomness-deficiency}
\delta (x  \mid  S) = \log |S| - K(x  \mid  S),
      \end{equation}
for $x \in S$, and $\infty$ otherwise.

{\bf ``Best Fit'' function:}
The {\em minimal randomness deficiency} function is
           \begin{equation}
\label{eq1}
\beta_x( \alpha) =
\min_{S} \{ \delta(x \mid  S): S \ni x, \;  K(S) \leq \alpha \},
            \end{equation}
where we set $\min \emptyset = \infty$.
The smaller $\delta(x  \mid  S)$ is, the more $x$ can be considered
as a {\em typical} member
of $S$. This means that a set $S$ for which $x$ incurs minimal
deficiency, in the model class of contemplated sets of given maximal
Kolmogorov complexity, is a ``best fitting'' model
for $x$ in that model class---a most likely explanation, and $\beta_x(\alpha)$
can be viewed as a {\em constrained best fit estimator}.
If the randomness deficiency is close to 0,
then are no simple special properties that
single it out from the majority of elements in $S$.
This is not just terminology: If $\delta (x  \mid  S)$ is small enough, then $x$
satisfies {\em all} properties of low Kolmogorov complexity
that hold with high probability for the elements of $S$. To be precise:
Consider strings of length $n$ and let $S$ be a subset of such
strings. A {\em property} $P$ represented by $S$ is a 
subset of $S$, and we say that
$x$ satisfies property $P$ if $x \in P$. Often, the cardinality
of a family of sets $\{S\}$ we consider depends
on the length $n$ of the strings in $S$. We discuss 
properties in terms of bounds $\delta (n) \leq \log |S|$. 
(The lemma below can also be formulated in terms of
probabilities instead of frequencies if we are talking
about a probabilistic ensemble $S$.)

\begin{lemma}
Let $S \subseteq \{0,1\}^n$.

(i) If $P$ is a property satisfied by all $x \in S$ with
$\delta(x  \mid  S) \le \delta (n)$,
then $P$ holds for a fraction of at
least $1-1/2^{\delta(n)}$ of the elements in $S$.

(ii) Let $n$ and $S$ be fixed, and let
$P$ be any
property
that holds for a fraction of at least
$1-1/2^{\delta (n)}$ of the
elements of $S$. 
There is a constant $c$, such that every such $P$ holds
simultaneously for every $x \in S$
with $\delta (x  \mid  S)\le\delta (n)-K(P \mid S) -c$.
\end{lemma}

\begin{proof}
(i) There are only  $\sum_{i=0}^{\log |S| - \delta (n)}2^i$
programs of length not greater than $\log |S| - \delta (n)$
and there are $|S|$ elements in $S$.

(ii) 
Suppose $P$ does not hold for an object $x \in S$
and the randomness deficiency satisfies
$\delta(x|S) \leq \delta (n) -K(P|S)-c$.
Then we can reconstruct $x$ from a description of $P$, which can use $S$,
and $x$'s index $j$ in an effective enumeration of all objects for
which $P$ doesn't hold. There are at
most $|S|/2^{ \delta (n)}$ such
objects by assumption, and therefore there are constants $c_1,c_2$ such that
\[ K(x|S) \leq  \log j+ c_1 \leq \log |S| - \delta ( n) + c_2. \]
Hence, by the assumption on the randomness deficiency of
$x$, we find $K(P|S) \leq c_2 -c$,
which contradicts the necesssary nonnegativity
of $K(P|S)$ if we choose $c > c_2$.
\end{proof}

\begin{example}
\rm
{\bf Lossy Compression}
 The function $\beta_x( \alpha)$ is relevant to lossy compression
(used, for instance, to compress images).
Assume we need to compress $x$ to $\alpha$ bits where $\alpha\ll K(x)$.
Of course this implies some loss of information present in $x$.
One way to select redundant information to discard is as follows:
Find a set $S\ni x$ with $K(S)\le\alpha$ and with small $\delta(x | S)$,
and consider a compressed version $S'$ of $S$.
To reconstruct an $x'$,
a decompresser uncompresses $S'$ to $S$ and selects
at random an element $x'$ of $S$. 
Since with high probability the randomness deficiency of $x'$ in $S$ is small,
$x'$ serves the purpose of the message $x$ as well as does $x$ itself.
Let us look at an example. To transmit a picture of
``rain'' through a channel with limited capacity $\alpha$,
one can transmit the indication that this is a picture of the rain and
the particular drops may be chosen by the receiver at random.
In this interpretation, $\beta_x(\alpha)$ indicates
how ``random'' or ``typical'' $x$ is with respect to the best model
at complexity level $\alpha$---and hence how ``indistinguishable'' from the
original $x$ the
randomly reconstructed $x'$ can be expected to be.
The relation of the structure function to lossy compression
and rate-distortion theory is the subject of an upcoming paper by the authors.
\end{example}

{\bf ``Structure'' function:}
The original Kolmogorov {\em structure} function 
\cite{Ko74a,Ko74b} for data $x$ is defined as
 \begin{equation}\label{eq2}
   h_{x}(\alpha) = \min_{S} \{\log | S| : S \ni x,\; K(S) \leq \alpha\},
\end{equation}
where $S \ni x$ is
a contemplated model for $x$, and $\alpha$ is a nonnegative
integer value bounding the complexity of the contemplated $S$'s.
Clearly, this function is
non-increasing and reaches $\log |\{x\}| = 0$
for $\alpha = K(x)+c_1$ where $c_1$ is the number of bits required
to change $x$ into $\{x\}$. The function can also
be viewed as a {\em constrained maximum likelihood (ML) estimator},
a viewpoint that is more evident for its version 
for probability models, Figure~\ref{figure.MLestimator} 
in Appendix~\ref{app.extension}.
For every $S\ni x$ we have
         \begin{equation}\label{eq.descr}
K(x)\leq K(S)+ \log |S| + O(1).
          \end{equation}
Indeed,
consider the following \emph{two-part code}
for $x$: the first part is
a shortest  self-delimiting program $p$ of $S$ and the second
part is
$\lceil\log|S|\rceil$ bit long index of $x$
in the lexicographical ordering of $S$.
Since $S$ determines $\log |S|$ this code is self-delimiting
and we obtain \eqref{eq.descr}
where the constant $O(1)$ is
the length of the program to reconstruct
$x$ from its two-part code.
We thus conclude that $K(x)\le\alpha+h_x(\alpha)+O(1)$, that is, the
function $h_x(\alpha)$
never decreases
more than a fixed independent constant below
the diagonal \emph{sufficiency line} $L$ defined by
$L(\alpha)+\alpha = K(x)$,
which is a lower bound on $h_x (\alpha)$
and is approached to within a constant distance by
the graph of $h_x$ for certain $\alpha$'s
(for instance, for $\alpha = K(x)+c_1$).
For these $\alpha$'s we
have
$\alpha + h_x (\alpha) = K(x)+O(1)$ and the associated model
(witness for
$h_x(\alpha)$) is called an {\em optimal set} for $x$,
and its description of $\leq \alpha$ bits is
called a {\em sufficient statistic}. If no confusion can result
we use these names interchangeably. The main properties 
of a sufficient statistic are the following: If $S$ is a sufficient
statistic for $x$, then $K(S)+\log |S| = K(x)+O(1)$. That is,
the two-part description of $x$ using the model $S$ and
as data-to-model code the index of $x$ in the enumeration
of $S$ in $\log |S|$ bits, is as concise as the shortest one-part
code of $x$ in $K(x)$ bits. Since now $K(x) \leq K(x,S) +O(1)
\leq K(S)+K(x|S)+O(1) \leq K(S)+\log|S|+O(1) \leq K(x)+O(1)$,
using straightforward inequalities (for example, given $S \ni x$, 
we can describe $x$ self-delimitingly in $\log |S|+O(1)$ bits)
and the sufficiency property,
we find that $K(x|S)=\log |S| +O(1)$. Therefore,
the randomness deficiency of $x$ in $S$ is constant,
$x$ is a typical
element for $S$, and $S$ is a model of best fit for $x$.
The data item $x$ can have randomness deficiency about 0, and
hence be a typical element 
for models $S$ that are not sufficient statistics. 
A sufficient statistic $S$ 
for $x$ has the additional property, apart from being a model
of best fit, that $K(x,S)=K(x)+O(1)$
and therefore by \eqref{eq.soi} we have $K(S|x^*)=O(1)$:
the sufficient statistic $S$ is a model of best fit
that is almost completely determined by $x$.
The sufficient
statistic associated with the least such $\alpha$ is called the
{\em minimal sufficient statistic}. For more details see
\cite{CT91,GTV01} and Section~\ref{sect.appl}.

{\bf ``Minimal Description Length'' function:}
The length of the minimal two-part code for $x$ consisting
of the model cost $K(S)$ and the
length of the index of $x$ in 
$S$, in the model class of sets of given maximal Kolmogorov complexity $\alpha$,
the complexity of $S$ upper bounded by $\alpha$, is given by
the {\em MDL} function or {\em constrained MDL estimator}:
  \begin{equation}\label{eq.3}
   \lambda_{x}(\alpha) =
\min_{S} \{\Lambda(S): S \ni x,\; K(S) \leq \alpha\},
  \end{equation}
where $\Lambda(S)=\log|S|+K(S) \ge K(x)-O(1)$ is
the total length of two-part code of $x$
with help of model $S$.
Apart from being convenient for the technical analysis
in this work, $\lambda_x (\alpha)$ is the
celebrated two-part Minimum Description Length code
length (Section~\ref{ex.MDL}) as a function of $\alpha$,
with the model class restricted to models
of code length at most $\alpha$.

\section{Overview of Results}
\subsection{Background and Related Work}
There is no written version, apart from the few lines 
which we reproduced in Section \ref{sect.trans},
of A.N. Kolmogorov's initial proposal  \cite{Ko74b,Ko74a} 
for a non-probabilistic approach to Statistics and Model Selection.
We thus have to rely on oral history, see Appendix~\ref{app.oral}.
There, we also describe an early 
independent related result of L.A. Levin \cite{Le02}.
Related work on so-called ``non-stochastic objects''
(where $h_x (\alpha)+\alpha$
drops to $K(x)$ only for large $\alpha$) is
\cite{Sh83,Vy87,Sh99,ViLi99,Vi01}.
In 1987, \cite{Vy87,Vy99}, V.V.
V'yugin established that,
for $\alpha = o(|x|)$,
 the randomness deficiency function $\beta_x(\alpha)$
can assume all possible shapes
(within the obvious constraints). 
In the survey \cite{CGG89} of Kolmogorov's work
in information theory, the authors preferred to mention
$\beta_x(\alpha)$, because it by definition optimizes ``best fit,''
rather than $h_x(\alpha)$ of which the usefulness
and meaningfulness was mysterious.
But Kolmogorov had
a seldom erring intuition: we will show that his original
proposal $h_x$ in the proper sense incorporates
all desirable properties of $\beta_x(\alpha)$, and in
fact is superior.
In \cite{Co85,CT91,CGG89} a notion of ``algorithmic sufficient statistics'',
derived from Kolmogorov's structure function,
is suggested as the algorithmic approach to
the probabilistic notion of sufficient statistic
\cite{Fi22,CT91} that is central in classical statistics.
The paper \cite{GTV01}
investigates the algorithmic notion in detail
and formally establishes such a relation.
The algorithmic (minimal) sufficient statistic
is related in \cite{ViLi99,GLV00} to the
``minimum description length'' principle \cite{Ri83,BRY,Wa87} in statistics
and inductive reasoning. Moreover, \cite{GTV01} observed that
$\beta_x (\alpha ) \leq h_x (\alpha) + \alpha - K(x) +O(1)$, establishing
a one-sided relation between (\ref{eq1}) and (\ref{eq2}), and the question
was raised whether the converse holds.

\subsection{This Work}
When we compare statistical hypotheses
$S_0$ and $S_1$ to explain data $x$ of length $n$, we should take into account
three parameters:
$K(S), K(x  \mid S)$, and $\log|S|$.
The first parameter is the {\em simplicity} of the
theory $S$ explaining the data.
The difference $\delta(x|S)=\log|S|-K(x  \mid S)$
(the randomness deficiency)
shows {\em how typical} the data is
with respect to $S$.
The sum $\Lambda(S)=K(S)+\log |S|$
tells us how {\em short
the two part code}
of the data using theory $S$ is, consisting
of the code for $S$ and a code for $x$ simply using the worst-case
number of bits possibly required to identify $x$ in 
the enumeration of $S$. This second part consists of the full-length index
ignoring savings in code length using possible non-typicality
of $x$ in $S$ (like being the first element in the enumeration of $S$).
We would like to define that $S_0$ is not worse than $S_1$
(as an explanation for $x$),
in symbols: $S_0 \le S_1$,  if
\begin{itemize}
\item
$K(S_0) \le K(S_1)$;
\item
$\delta(x|S_0) \le \delta(x|S_1)$; and
\item
$\Lambda(S_0) \le \Lambda(S_1)$.
\end{itemize}
To be sure,  this is not equivalent to saying that
$K(S_0) \le K(S_1),
\delta(x|S_0) \le \delta(x|S_1),
\log|S_0| \le \log|S_1|$.
(The latter relation is stronger in that it implies 
$S_0 \le S_1$ but not vice versa.)
The algorithmic statistical properties of a data string $x$ are
fully represented by
the set $A_x$ of all
triples
$\pair{ K(S), \delta(x|S), \Lambda(S) }$
such that $S \ni x$, together with a component wise 
order relation $\le$ on the elements those triples.
The complete characterization of 
how this set may look like
(with $O(\log n)$-accuracy) is now known in the following sense.

Our results (Theorems~\ref{th14}, \ref{th13}, \ref{thsxa})
describe completely
(with $O(\log n)$-accuracy) possible shapes of the closely related set
$B_x$ consisting of all triples $\pair{\alpha,\beta,\lambda}$ such that
there is a set $S\ni x$ with
$K(S)\le\alpha$,
$\delta(x \mid S)\le\beta$,
$\Lambda(S)\le\lambda$. That is, $A_x \subseteq B_x$ and
$A_x$ and $B_x$
have the same minimal triples. Hence,
we can informally say that our results describe completely
possible shapes of the set of
triples
$\pair{ K(S), \delta(x|S), \Lambda(S) }$
for non-improvable hypotheses $S$ explaining $x$.
For example up to $O(\log n)$
accuracy, and denoting $k=K(x)$ and $n=|x|$:

   (i) For every minimal triple $(\alpha,\beta,\gamma)$ in $B_x$
   we have $0 \le \alpha \le k$, $0 \le \beta, \beta+k = \gamma \leq n$.

   (ii) There is a triple of the form $(\alpha_0,0,k)$ in $B_x$
   (the minimal such $\alpha_0$ is the complexity
   of the minimal sufficient statistic for $x$). This property allows
   us to recover the complexity $k$ of $x$ from $B_x$.

   (iii) There is a triple of the form $(0,\lambda_0-k,\lambda_0)$
   in $B_x$ with $\lambda_0 \le n$.

Previously,
a limited characterization was obtained by
V'yugin~\cite{Vy87,Vy99} for the possible shapes of the projection of $B_x$
on $\alpha,\beta$-coordinates but only for the case when
$\alpha=o(K(x))$.
Our results
describe possible shapes of the entire set
$B_x$ for
the full domain of $\alpha$ (with $O(\log n)$-accuracy).
   Namely, let $f$ be a non-increasing integer valued function
   such that $f(0)\le n$,  $f(i)=k$ for all $i\ge k$ and
           $$
   \tilde B_{f}=\{\pair{\alpha,\beta,\lambda}\mid 
   0\le\alpha,\
   f(\alpha)\le\lambda,\
   f(\alpha)-k\le\beta\}.
           $$
For every $x$ of length $n$ and complexity $k$ there is
$f$ such that
	\begin{equation}\label{eq-final}
\tilde B_{f}+ u \subset B_x \subset \tilde B_{f} - u
	\end{equation}
where $u=\pair{c\log n,c\log n,c\log n}$ for some universal constant $c$.
Conversely, for every $k\le n$ and every such $f$ there is
$x$ of length $n$
such that~\eqref{eq-final} holds for $u=\pair{c\log K(n,f,k),c\log
K(n,f,k),c\log K(n,f,k)}$ . Our results imply that the set $B_x$
is not computable given $x, k$ but is computable given $x, k$ and
   $\alpha_0$, the complexity of minimal sufficient statistic.

   \begin{remark}
\rm
   There is also the fourth important parameter, $K(S \mid x^*)$ reflecting
   the {\em determinacy of model} $S$ by the data $x$.
   However, the equality $\log |S|+K(S)-K(x)  = K(S \mid x^*)+ \delta (x \mid S)
   +O(1)$ shows that this parameter can be expressed in $\alpha, \beta, h$.
   The main result (\ref{eq.eq}) establishes
   that $K(S \mid x^*)$ is logarithmic
   for {\em every} set $S$ witnessing $h_x (\alpha)$. This also
   shows that there are at most polynomially many such sets.
\end{remark}

\subsection{Technical Details}
The results are obtained by analysis of the relations between
the structure functions.
The most fundamental result in this paper
is the equality
         \begin{equation}\label{eq.eq}
\beta_x (\alpha )  = h_x (\alpha) + \alpha - K(x) = \lambda_x (\alpha)
- K(x)
         \end{equation}
which holds within additive terms, that are logarithmic in the length of the 
string $x$, in argument and value.
Every set $S$ that witnesses the value $h_x (\alpha )$
(or $\lambda_x(\alpha)$),
also witnesses the value $\beta_x (\alpha)$ (but not vice versa).
It is easy to see that $h_x (\alpha)$ and $\lambda_x(\alpha)$
are
upper semi-computable (Definition~\ref{def.semi});
but we show that $\beta_x (\alpha)$ is neither upper nor lower semi-computable.
A priori
there is no reason to suppose that
a set that witnesses $h_x (\alpha)$
(or $\lambda_x(\alpha)$) also witnesses $\beta_x (\alpha)$,
for {\em every} $\alpha$.
But the fact that they do, vindicates
Kolmogorov's original proposal and establishes $h_x$'s pre-eminence
over $\beta_x$.
The result can be taken as a foundation and justification
of common statistical principles in model
selection such as maximum likelihood
or MDL (\cite{Ri83,BRY} and our Sections~\ref{ex.MDL} and~\ref{ex.ML}).
We have also addressed 
the fine structure of the shape of $h_x$ (especially
for $\alpha$ below the minimal sufficient statistic complexity)
and a uniform (noncomputable) construction for the structure functions.

The possible (coarse) shapes of the functions
$\lambda_x,h_x$ and $\beta_x$ are examined
in Section~\ref{sect.coarse}. Roughly stated:
The structure functions $\lambda_x,h_x$ and $\beta_x$ can assume all
possible shapes over their full domain of definition (up to
additive logarithmic precision in both argument and value).
As a consequence, so-called ``non-stochastic'' strings $x$ 
for which $h_x(\alpha)+\alpha$
stabilize on $K(x)$ for large $\alpha$ are common.
This
improves and extends V'yugin's result \cite{Vy87,Vy99} above;
it also improves the independent
related result of L.A. Levin \cite{Le02} in Appendix~\ref{app.oral}; and,
applied to ``snooping curves'' extends a recent result
of V'yugin, \cite{Vy01}, in Section~\ref{ex.snoop}.
    The fact that  
    $\lambda_x$ can assume all
    possible shapes over its full domain of definition
    establishes
    the significance of \eqref{eq.eq}, since it shows that $\lambda_x (\alpha)
    \gg K(x)$ indeed happens for some $x, \alpha$ pairs. 
    In that case the more or less
    easy fact that $\beta_x(\alpha)=0$ for $\lambda_x(\alpha)=K(x)$ is not
    applicable, and {\em a priori} there is no reason
    for \eqref{eq.eq}:
    Why should minimizing a set containing
    $x$ plus the set's description length
also minimize $x$'s randomness deficiency in the set? But
\eqref{eq.eq} shows that it does!
We determined the (fine) details of the function shapes
in Section~\ref{sect.fine}.
(Non-)computability properties are examined in Section~\ref{sect.comput},
 incidentally proving a to our knowledge first natural example,
$\beta_x$, of a function that
is not semi-computable but computable with an oracle for the halting problem.
In Section~\ref{sect.real}, we exhibit a uniform
construction for sets
realizing $h_x (\alpha)$
for all $\alpha$.

\subsection{Probability Models}
Following Kolmogorov we analyzed a canonical setting
where the models are finite sets. As Kolmogorov himself pointed
out, this is no real restriction: 
the finite sets model class is equivalent, up to a logarithmic additive
term, to the model class of probability density functions, as
studied in \cite{Sh83,GTV01}, and
the model class of total recursive functions, as studied in \cite{Vi01},
see Appendix~\ref{app.extension}.

\subsection{All Stochastic Properties of the Data}
The result (\ref{eq.eq}) shows that
the function $h_x(\alpha)$ yields all stochastic properties of data $x$
in the following sense:
for every $\alpha$ the class of models of maximal complexity 
$\alpha$ has a best model with goodness-of-fit determined by the
randomness deficiency $\beta_x(\alpha) =
h_x(\alpha)+\alpha - K(x)$---the equality being taken up to logarithmic
precision. For example, for some
value $\alpha_0$ the minimal randomness deficiency $\beta_x(\alpha)$
may be quite large for $\alpha < \alpha_0$ (so the best model in that
class has poor fit), but an infinitessimal increase in model complexity
may cause $\beta_x(\alpha)$ to drop to zero (and hence the marginally
increased model class now has a model of perfect fit), see
Figure~\ref{figure.estimator}. Indeed, the
structure function quantifies the best possible fit for a model in classes of
every complexity. 

\subsection{Used Mathematics}
Kolmogorov's proposal for a nonprobabilistic statistic
is combinatorial and algorithmic,
rather than probabilistic.
Similar to other recent directions in information theory and statistics,
this involves notions and proof techniques 
from computer science theory, rather
than from probability theory. But the contents matter
and results are about traditional statistic- and information theory
notions like model selection, information and compression;
consequently the treatment straddles fields that are not traditionally
intertwined. 
For convenience of the reader who is unfamiliar with algorithmical
notions and methods we have taken pains to provide intuitive
explanations and interpretations. Moreover, we have delegated almost all proofs
to Appendix~\ref{app.proof}, and all precise formulations and proofs
of the (non)computability and (non)approximability of the structure functions
to Appendix~\ref{app.comp}.
\section{Coarse Structure}
\label{sect.coarse}
In classical statistics, unconstrained maximal likelihood is known
to perform badly for model selection, because
it tends to want the most complex models possible. A precise quantification
and explanation of this phenomenon, in the complexity-constrained
model class setting, is given in this section.
It is easy to see that unconstrained maximization
will result in the singleton set model $\{x\}$ of complexity
about $K(x)$.
We will show that the structure function
$h_x(\alpha)$ tells us all stochastic properties of data $x$.
From complexity 0 up to the complexity where the
graph hits the sufficiency line, the best fitting models
do not represent all meaningful properties of $x$. The distance
between $h_x (\alpha)$ and the sufficiency line $L(\alpha)=K(x)-\alpha$,
is a measure, expressed by $\beta_x(\alpha)$, of how far the best 
fitting model at complexity $\alpha$
falls short of a sufficient fitting model. 
The least complex sufficient fitting model,
the minimal sufficient statistic, 
occurs at 
complexity level $\alpha_0$ where $h_x$ hits
the sufficiency line. There, $h_x(\alpha_0)+\alpha_0 = K(x)$.
The minimal sufficient statistic model expresses all meaningful
information in $x$, and its complexity is the number of bits
of meaningful information in the data $x$. 
The remainder $h_x(\alpha_0)$ bits of the $K(x)$ bits of  information 
in data $x$ is the ``noise,'' the meaningless randomness, contained in 
the data.
When we consider the
function $h_x$ at still higher complexity levels $\alpha > \alpha_0$, 
the function $h_x(\alpha)$ hugs the sufficiency line  $L(\alpha)=K(x)-\alpha$,
which means that $h_x(\alpha)+\alpha$ stays constant at $K(x)$.
The best fitting models at these complexities start to model
more and more noise, $h_x(\alpha)-h_x(\alpha_0) = \alpha_0 - \alpha$ bits, 
in the data $x$: 
the added complexity $\alpha_0 - \alpha$ 
in the sufficient statistic model 
at complexity level $\alpha$ over
that of the minimal sufficient statistic at complexity level $\alpha_0$ 
is completely used to
model increasing part of the noise in the data. 
The worst overfitting occurs when we arrive at complexity $K(x)$,
at which point we model all noise in the data apart from the meaningful
information. Thus, our approach makes the fitting process
of {\em constrained maximum likelihood}, first underfitting at
low complexity levels of the models considered,
then the complexity level of optimal fit (the minimal
sufficient statistic), and subsequently the overfitting at higher
levels of complexity of models, completely and formally
explicit in terms of fixed data and individual models.

\subsection{All Shapes are Possible}
\begin{figure}
\begin{center}
\epsfxsize=8cm
\epsfxsize=8cm \epsfbox{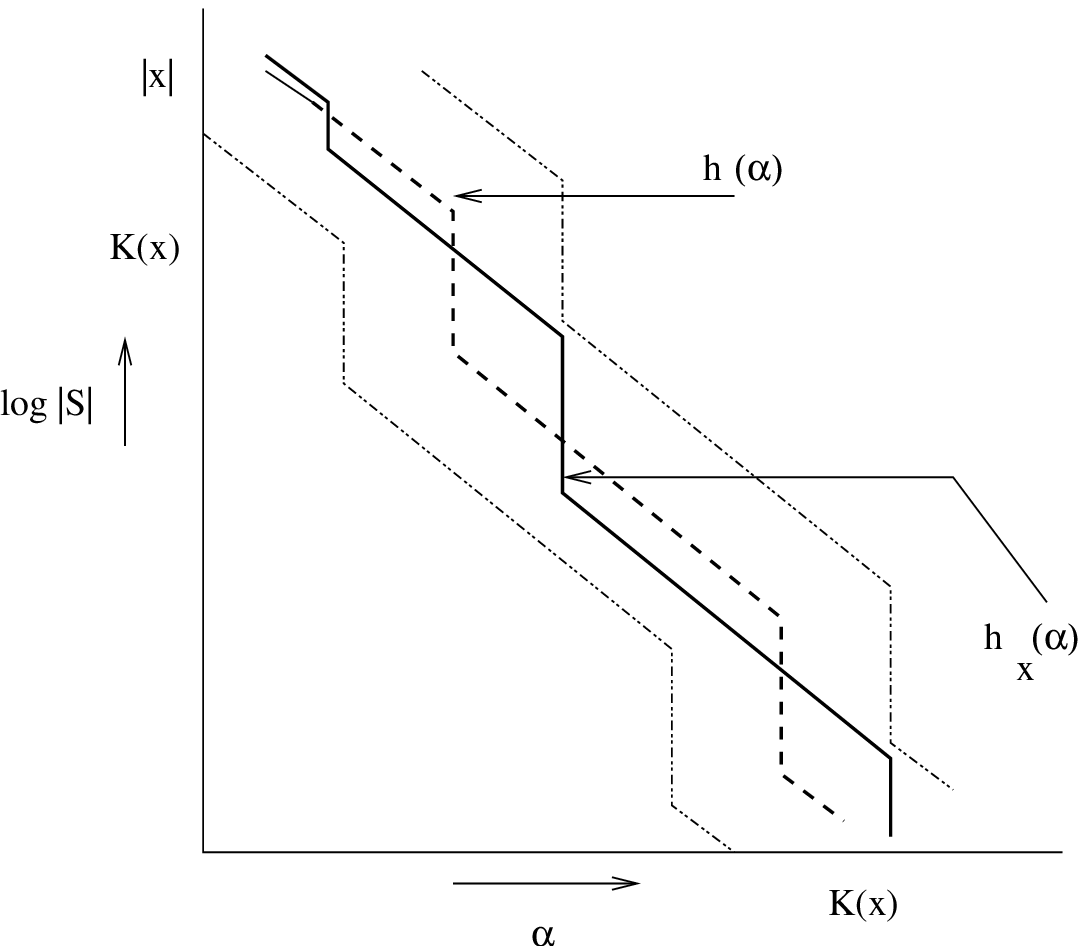}
\end{center}
\caption{Structure function $h_x( \alpha )$ 
in strip determined by $h( \alpha )$,
that is, $\close{h_x( \alpha )}{h( \alpha )}$.}
\label{figure.close}
\end{figure}

Let $\beta_x (\alpha)$ be defined as in
(\ref{eq1}) and  
$h_x (\alpha)$ be defined as in
(\ref{eq2}). Both functions are 0 ($\beta_x (\alpha)$ may be $-O(1)$)
 for all $\alpha \geq K(x)+c_0$
where $c_0$ is a constant.
We {\em represent} the coarse shape of these functions for different $x$  
by functions characteristic of that shape.
Informally, $g$ represents $f$ means that the graph of $f$ is contained
in a strip of logarithmic (in the length $n$ of $x$) width centered
on the graph of $g$, Figure~\ref{figure.close}. 

{\em Intuition: $f$ follows $g$ up to a prescribed precision.}

For formal statements
we rely on the notion in Definition~\ref{def.close}.
Informally, we obtain the following results
($x$ is of length $n$ and complexity $K(x)=k$):

\begin{itemize}
\item 
Every non-increasing function  $\beta$
represents  $\beta_x$ for
some $x$, and for every $x$
the function $\beta_x$ is represented by some $\beta$,
provided 
$\beta(k)= 0$, $\beta(0) \leq n-k$.
\item 
Every function $h$, 
with non-increasing $h(\alpha)+\alpha$,
represents $h_x$ for
some $x$, and for every $x$ the function $h_x$ is represented
by some $h$ as above, provided 
$h(k) = 0$, $h(0) \leq n$ (and by the non-increasing property $h(0) \geq k$).
\item
$h_x(\alpha)+\alpha$ represents $\beta_x(\alpha)+k$, and conversely,
for every $x$.
\item For every $x$ and $\alpha$, 
every minimal size set $S\ni x$ of complexity 
at most $\alpha' = \alpha+O(\log n)$, has randomness deficiency
$\beta_x (\alpha') \leq \delta(x \mid S) \leq \beta_x(\alpha) + O(\log n)$.
\end{itemize}

To provide  precise statements we need a definition.

	\begin{definition}\rm
\label{def.close}
Let $f,g$ be functions defined on $\{0,1,\dots,k\}$ 
with values in $\N\cup\{\infty\}$. We say that   
$f$ is $(\eps(i),\delta(i))$-close to $g$ (in symbols: $\close fg$) if
	\begin{align*}
f(i)&\ge \min\{g(j)\colon j\in[\eps(0),k],\ |j-i|\le\eps(i)\}-\delta(i)\},\\
f(i)&\le \max\{g(j)\colon  j\in[\eps(0),k],\ |j-i|\le\eps(i)\}+\delta(i)\}
	\end{align*}
for every $i\in[\eps(0),k]$. If $\close fg$ and $\close gf$
we write $\dclose fg$.
	\end{definition}

Here $\eps(i), \delta(i)$ are small values like $O(\log n)$ when we consider
data $x$ of length $n$.
Note that this definition is not symmetric and allows 
$f(i)$ to have arbitrary values for $i\in[0,\eps(0))$.
However, it is transitive in the following sense: 
if $f$ is $(\eps_1(i),\delta_1(i))$-close to
$g$ and $g$ is $(\eps_2(i),\delta_2(i))$-close to $h$ then $f$ is
$(\eps_1(i)+\eps_2(i), \delta_1(i) + \delta_2(i))$-close to $h$.
If $\close fg$ and $g$ is linear continuous, meaning that 
$|g(i)-g(j)|\le c|i-j|$ for some constant
$c$, then the difference between $f(i)$ and  $g(i)$ is bounded by  
$c\eps(i)+\delta(i)$ for every $\eps(0)\le i\le k$.

This notion of closeness, if applied unrestricted,
is not always meaningful. For example,
take as $g$ the function taking value $n$ for all
even $i\in[0,k]$ and 0 for all odd $i\in[0,k]$.
Then for {\em every} function $f$
on $[0,k]$ with $f(i)\in[0,n]$ we have $\close fg$
for $\eps=1$, $\delta=0$.
But if
$\close fg$ and $g$ is {\em non-increasing} then $g$
indeed gives much information about $f$.

It is instructive to consider the following example.
Let $g(i)$ be equal to $2k-i$ for $i=0,1,\dots, \frac{k}{2}-1$ and 
to $k-i$ for $i= \frac{k}{2},\dots,k$.	Let $\eps(i),\delta(i)$ be constant.
Then a function $\close fg$ may take every value
for $i\in[0,\eps)$, every value in $[2k-i-2\delta,2k-i+2\delta]$ for
$i\in[\eps,\frac{k}{2}-\delta]$,
every value in $[k-i-\delta,2k-i+\delta]$ for
$i\in(\frac{k}{2}-\delta,\frac{k}{2}+\delta]$, and every value in $[k-i-2\delta,k-i+2\delta]$ for
$i\in(\frac{k}{2}+\delta,k]$ (see Figure~\ref{figure.close}). 
Thus the point $\frac{k}{2}$ of discontinuity of $g$ 
gives an interval of size $2\delta$ of large ambiguity of $f$.
Loosely speaking the graph of $f$ can be any function
contained in the strip of radius
$2\delta$ whose middle line is the graph of $g$.   
For technical reasons 
it is convenient to use, in place of $h_x$, the MDL
function $\lambda_x$ (\ref{eq.3}).
The definition of $\lambda_x$ immediately implies the following properties:
$\lambda_{x}(\alpha)$ is non-increasing,
$\lambda_{x}(\alpha)\ge K(x)-O(1)$ for all $\alpha$.

The next lemma shows that properties of $\lambda_x$ translate
directly into properties of $h_x$ since
$h_x (\alpha)$ is always ``close'' to
$\lambda_x(\alpha) - \alpha$.

      \begin{lemma}\label{lem.equiv}
For every $x$ we have
		 $
\lambda_x(\alpha)\le h_x(\alpha)+\alpha\le
\lambda_x(\alpha)+K(\alpha)+O(1)
		 $
for all $\alpha$. Hence
$\dclose{\lambda_x(\alpha)}{h_x(\alpha)+\alpha}$ for $\eps=0$,
$\delta=K(\alpha)+O(1)$.
      \end{lemma}

{\em Intuition: 
The  functions $h_x (\alpha) + \alpha$ (the ML code length plus 
the model complexity) 
and $\lambda_x(\alpha)$ (the MDL code length) are essentially
the same function.
}

\begin{remark}
\rm
The lemma implies that the {\em same set} witnessing $h_x (\alpha)$ also
witnesses $\lambda_x(\alpha)$
up to an additive term of $K(\alpha)$.
The converse is only true for the smallest cardinality set witnessing
$\lambda_x (\alpha)$. Without this restriction
a counter example is:  for random $x\in\{0,1\}^n$ the set
$S=\{0,1\}^n$ witnesses
$\lambda_x(\frac{n}{2})=n+O(K(n))$
but does not witness
$h_x(\frac{n}{2})= \frac{n}{2}+O(K(n))$. (If $\lambda_x (\alpha) = K(x)$,
 then every
set of complexity $\alpha' \leq \alpha$
witnessing $\lambda_x(\alpha') = K(x)$ also witnesses
$\lambda_x (\alpha) = K(x)$.)
\end{remark}

The next two theorems state the main results of this work in a precise form. 
By $K(i,n,\lambda)$ we mean the minimum length of a program that
outputs $n,i$, and computes
$\lambda(j)$ given any $j$ in the domain of $\lambda$. We first
analyze the possible shapes of the structure functions.

	   \begin{theorem}\label{th14}
(i) For every $n$ and every string $x$ of length $n$ and complexity 
$k$ there is an
integer valued non-increasing function $\lambda$ defined on $[0,k]$
such that
$\lambda(0)\le n$, $\lambda(k)=k$ and
$\close{\lambda_x}\lambda$ for $\eps=\delta=K(n)+O(1)$.

(ii) Conversely, for every $n$ and non-increasing integer valued
function $\lambda$ whose domain includes $[0,k]$
and such that
$\lambda(0)\le n$ and  $\lambda(k)=k$,
there is $x$ of length $n$ and complexity 
$k\pm (K(k,n,\lambda)+O(1))$ such that $\close{\lambda_x}\lambda$
for
$\eps = \delta=K(i,n,\lambda)+O(1)$.
	   \end{theorem}

{\em Intuition:
The MDL code length $\lambda_x$, and therefore
by Lemma~\ref{lem.equiv} also the original structure function $h_x$,
can assume essentially every possible shape as a function of the
contemplated maximal model complexity.
}

              \begin{remark}
    \label{rem.uncondh}
    \rm
    The theorem implies that for every function $h(i)$ defined on $[0,k]$
    such that the
    function $\lambda (i)=h(i)+i$ satisfies the conditions of item
    (ii) there is an $x$ such that $\close{h_x(i)}{ h(i)}$ with
    $\eps=\delta=O(K(i,n,h))$.
              \end{remark}

\begin{remark}
\label{rem.condh}
\rm
The proof of the theorem shows that for every function $\lambda (i)$
satisfying the conditions of item (ii)
there is $x$ such that
$\close{\lambda_x (i\mid n,\lambda)}{\lambda(i)}$ with	$\eps=\delta=K(i)+O(1)$
where the conditional structure function $\lambda_x(i \mid y) =
\min_S \{ K(S \mid y)+\log |S| : S \ni x , \; K(S \mid y ) \leq i \}$.
Consequently, for
every function $h(i)$ such that the function $\lambda (i)=h(i)+i$
satisfies the conditions of item (ii)
there is an $x$ such that $\close{h_x(i \mid n,h)}{ h(i)}$
with $\eps=\delta=O(K(i))$
where the conditional structure function $h_x(i \mid y) =
\min_S \{ \log |S| : S \ni x , \; K(S \mid y ) \leq i \}$.
\end{remark}

\begin{remark}
\label{rem.2}
\rm
In the proof of Item (ii) of the theorem we can consider
every finite set $U$ with $|U| \geq 2^n$ in place
of the set $A$ of all strings
of length $n$. Then we obtain a string $x \in U$ such that
$\close{\lambda_x}{\lambda}$ with $\eps(i)=\delta (i) = K(i,U,\lambda)$.
\end{remark}

\subsection{Selection of Best Fitting Model}
Recall that in classical statistics a major issue is whether a given 
model selection method works well if the ``right'' model
is in the contemplated model class, and what model the method selects
if the ``right'' model is outside the model class. 
We have argued earlier that the best we can do is to look
for the ``best fitting''
model. But both ``best fitting'' and ``best fitting in a constrained
model class'' are 
impossible to express classically for individual models and data. 
Instead, one focusses on probabilistic definitions and analysis.
It is precisely these issues
that can be handled in the Kolmogorov complexity setting.

For the complexity levels $\alpha$ at which $h_x(\alpha)$ 
coincides with the diagonal sufficiency line $L(\alpha) = K(x)-\alpha$,
the model class contains a ``sufficient'' (the ``best fitting'') model.

For the complexity levels $\alpha$ at which $h_x(\alpha)$ is
above the sufficiency line, the model class does not contain
a ``sufficient'' model. However,
our results say that $h_x(\alpha) - L(\alpha)$ equals
the minimal randomness deficiency that can be achieved
by a model of complexity $\leq \alpha$, and hence quantifies
rigorously the properties of the data $x$ such a model can represent,
that is, the level of ``fitness'' of the best model in the class.

Semi-computing $h_x(\alpha)$ from above, together with
the model wittnessing this value, automatically 
yields the objectively most fitting model in the class,
that is, the model that is closest to the ``true'' model
according to an objective measure
of representing most properties of data $x$.

The following central result of this paper shows that the
$\lambda_x$ (equivalently $h_x$, by Lemma~\ref{lem.equiv}) and $\beta_x$ 
can be expressed in one another but for
a logarithmic additive error.

	   \begin{theorem}\label{th13}
For every $x$ of length $n$ and complexity $k$
it holds $\dclose{\beta_x(\alpha)+k}{\lambda_x(\alpha)}$
for  $\eps=\delta=O(\log n)$.
	   \end{theorem}

{\em Intuition:
A model achieving the MDL code length $\lambda_x (\alpha)$, or
the ML code length $h_x (\alpha)$, essentially achieves the best possible fit
$\beta_x (\alpha)$.
}

		\begin{corollary}\label{cor.beta}
For every $x$ of length $n$ and complexity $k\le n$ 
there is a non-increasing function $\beta$ 
such that $\beta(0)\le n-k$, $\beta(k)=0$ and 
$\close{\beta_x}{\beta}$ for  $\eps,\delta=O(\log n)$.
Conversely, for every non-increasing function $\beta$  
such that 
$\beta(0)\le n-k$, $\beta(k)=0$ there is $x$ of length $n$ and complexity 
$k\pm \delta$ such that 
$\close{\beta_x}{\beta}$ for $\eps=\delta=O(\log n)+K(\beta)$.
		\end{corollary}
		
		\begin{proof}
The first part is more or less immediate. Or
use the first part of Theorem~\ref{th14} and then let 
$\beta(i)=\lambda(i)-k$.
To prove the second part, use the second part of Theorem~\ref{th13},
and the second part of Theorem~\ref{th14} with
$\lambda(i)=\beta(i)+k$.
		\end{proof}

\begin{remark}\label{rem.hbeta}
\rm
From the proof of Theorem~\ref{th13} we see that
for every finite set $S\ni x$, of complexity at most
$\alpha+O(\log n)$ and  minimizing $\Lambda(S)$,
we have
$\delta(x \mid S)\le \beta_x(\alpha)+O(\log n)$.
Ignoring $O(\log n)$ terms, at every complexity
level, every best hypothesis
at this level with respect to $\Lambda(S)$ is
also a best one with respect to
typicality. This explains why it is worthwhile to
find shortest two-part descriptions
for given data $x$: this is the single known
way to find an $S\ni x$ with respect to
which $x$ is as typical as possible at that complexity level.
Note that the set $\{\pair{x,S,\beta}\mid x\in S,\ \delta(x \mid
S)<\beta\}$ is not enumerable so we are not able to generate such
$S$'s directly (Section~\ref{sect.comput}).

The converse is not true: not every hypothesis, consisting of
a finite set, witnessing
$\beta_x(\alpha)$ also witnesses $\lambda_x(\alpha)$ or $h_x(\alpha)$.
For example, let $x$ be a
string of length $n$ with $K(x) \geq n$. Let $S_1=
\{0,1\}^n\cup\{y\}$ where $y$ is a string of length $\frac{n}{2}$ such
that $K(x,y)\ge \frac{3n}{2}$
and let $S_2= \{0,1\}^n$.
Then
both $S_1,S_2$ witness
$\beta_x(\frac{n}{2}+O(\log n))=O(1)$
but
$\Lambda(S_1)=\frac{3n}{2}+O(\log n)\gg
\lambda_x(\frac{n}{2}+O(\log n))=n+O(\log
n)$
while $\log |S_2| = n \gg h_x (\frac{n}{2}+O(\log n)) = \frac{n}{2}+ O(\log n)$.
\end{remark}

However, for every
$\alpha$ such that $\lambda_x(i)$ decreases when
$i \rightarrow \alpha$ with $i \leq \alpha$, a witness set for
$\beta_x (\alpha)$ is also a witness set for
$\lambda_x (\alpha)$ and $h_x(\alpha)$.
We will call such $\alpha$ {\em critical} (with respect to $x$): these are the
model complexities at which the two-part MDL code-length decreases,
while it is stable
in between such critical points.
The next theorem shows, for critical $\alpha$,
that
for every $A\ni x$ with
$K(A)\approx\alpha$ and $\delta(x \mid A)\approx\beta_x(\alpha)$,
we have $\log|A|\approx h_x(\alpha)$ and
$\Lambda(A)\approx \lambda_x(\alpha)$.
More specifically, if
$K(A)\approx\alpha$ and $\delta(x \mid A)\approx\beta_x(\alpha)$
but $\Lambda(A)\gg \lambda_x(\alpha)$ or
$\log|A| \gg h_x(\alpha)$ then
there is $S\ni x$ with
$K(S)\ll\alpha$ and
$\Lambda(S)\approx \lambda_x(\alpha)$.

       \begin{theorem}\label{thsxa}
For all $A\ni x$ there is $S\ni x$ such that
$\Lambda(S)\le \lambda_x(\alpha)+(\delta(x|A)-\beta_x(\alpha))$,
$K(S)\le K(A)+(\lambda_x(\alpha)-\Lambda(A))+
(\delta(x|A)-\beta_x(\alpha))$,
and
$K(S)\le \alpha+(h_x(\alpha)-\log|A|)+(\delta(x|A)-\beta_x(\alpha))$
where all inequalities hold up to $O(\log \Lambda(A))$
additive term.
        \end{theorem}

{\em Intuition:
Although models of best fit
(witnessing $\beta_x (\alpha)$) do not necessarily
achieve the MDL code length $\lambda_x (\alpha)$ or 
the ML code length $h_x (\alpha)$, they
do so at the model complexities where the MDL code length
decreases, and, equivalently, the ML code length
decreases at a slope of more than $-1$.
}

\subsection{Invariance under Recoding of Data}
\label{ex.recoding}
In what sense is the structure function invariant
under recoding of the data? Osamu Watanabe suggested the example
of replacing the
data $x$ by a shortest program $x^*$ for it.
Since $x^*$ is incompressible
it is a typical element of the set of all strings of length $|x^*|=K(x)$,
and hence $h_{x^*} (\alpha)$ drops to the sufficiency line
$L(\alpha) = K(x)-\alpha$ already for
some $\alpha \leq K(K(x))$, so almost immediately (and it
stays within logarithmic distance of that line henceforth).
That is,
$h_{x^*} (\alpha) = K(x)-\alpha$ up to
logarithmic additive terms in argument and value, 
irrespective of the (possibly quite different)
shape of $h_x$. Since the Kolmogorov complexity function
$K(x)=|x^*|$ is not recursive, \cite{Ko65},
the recoding function $f(x) = x^*$ is also not recursive.
Moreover,  while $f$ is one-one and total
it is not onto. 
But it is the
partiality of the inverse function (not all strings are shortest
programs) that causes the collapse of the structure function.
If one restricts the finite sets containing $x^*$ to be subsets of
$\{y^*: y \in \{0,1\}^*\}$, then the resulting
structure function $h_{x^*}$ is within a logarithmic strip around $h_x$.
However, the structure function
is invariant under ``proper'' recoding of the data. 

\begin{lemma}
Let $f$ be a recursive permutation of the set of finite binary strings
(one-one, total, and onto). Then,
$\close{h_{f(x)}}{h_x}$ for  $\eps,\delta=K(f)+O(1)$.
\end{lemma}
\begin{proof}
Let $S \ni x$ be a witness of $h_x(\alpha)$. Then,
$S_f = \{f(y): y \in S\}$ satisfies $K(S_f) \leq \alpha + K(f)+O(1)$
and $|S_f|=|S|$. Hence, $h_{f(x)} (\alpha + K(f)+O(1)) \leq h_x(\alpha)$.
Let $R \ni f(x)$ be a witness of $h_{f(x)} (\alpha)$. Then,
$R_{f^{-1}} = \{f^{-1} (y): y \in R\}$ satisfies
$K(R_{f^{-1}}) \leq \alpha + K(f)+O(1)$ and $|R_{f^{-1}}|=|R|$.
Hence, $h_{x} (\alpha + K(f)+O(1)) \leq h_{f(x)}(\alpha)$ (since 
$K(f^{-1}) = K(f)+O(1)$).
\end{proof}

\subsection{Reach of Results}
In Kolmogorov's initial proposal, as in this work, models are
finite sets of finite binary strings, and the data is one of the
strings (all discrete data can be binary encoded). The
restriction to finite set models is just a matter of convenience:
the main results generalize to the case where the models are
arbitrary computable probability density functions, 
\cite{Sh83,BC91,Sh99,GTV01},
and to the model class consisting of arbitrary total recursive
functions, \cite{Vi01}. We summarize the proofs of this below.
Since our results
hold only within
additive precision that is logarithmic in the binary length of the data, 
and the equivalences between the model classes
hold up to the same precision, the results hold equally for the
more general model classes.

The generality of the results are at the same time a restriction.
In classical statistics one is commonly interested in model classes
that are partially poorer and partially richer than the ones we consider.
For example, the class of Bernoulli processes, or $k$-state Markov
chains, is poorer than the class of computable probability density functions
of moderate maximal Kolmogorov complexity $\alpha$,  
in that the latter may contain
functions that require far more complex computations than the rigid
syntax of the former classes allows. Indeed, the class of computable
probability density functions of even moderate complexity allows
implementation of a function mimicking a universal Turing machine computation.
On the other hand, even the lowly Bernoulli process can be equipped
with a noncomputable real bias in $(0,1)$, and hence the generated
probability density function over $n$ trials is not a computable function.
This incomparability of the here studied algorithmic model classes, and
the traditionally studied statistical model classes, means that the
current results cannot be directly transplanted to the traditional setting.
They should be regarded as pristine truths that hold in a
platonic world that can be used as guideline to develop analogues
in model classes that are of more traditional concern, as in
\cite{Ri02}. The questions to be addressed are: Can these platonic 
truths say anything usable? If we restrict ourselves to
statistical model classes, how far from optimal are we?
Note that in themselves the finite set models are not really
that far from classical statistical models.

\section{Prediction and Model Selection}\label{sect.appl}

\subsection{Best Prediction Strategy}
\label{ex.snoop}
In~\cite{Vy01} the notion of a {\em snooping curve} $L_x(\alpha)$
of $x$ was introduced,
expressing the minimal logarithmic loss in predicting the
consecutive elements of a given individual string $x$, 
in each prediction using the preceding sequence of elements,
by the best prediction strategy of complexity at most $\alpha$. 

{\em Intuition: The snooping curve quantifies the quality
of the best predictor for a given sequence at every
possible predictor-complexity.
}

Formally,
$L_{x}(\alpha)=
\min \limits_{K(P)\le\alpha}\Loss_{P}(x)$.
The minimum is taken over all prediction strategies $P$ of
complexity at most $\alpha$.
A prediction strategy $P$ is a mapping
from the set of strings of length less than $|x|$
into the set of rational numbers in the segment $[0,1]$.
The value $P(x_1\dots x_i)$ is regarded as
our belief (or probability) that $x_{i+1}=1$
after we have observed $x_1,\dots,x_i$.
If the actual bit $x_{i+1}$ is 1 the strategy
suffers the loss $-\log p$ otherwise $-\log(1-p)$.
The strategy is a finite object and $K(P)$ may
by defined as the complexity of this object,
or as the minimum
size of a program that identifies $n=|x|$ and given $y$ finds
$P(y)$.
The notation $\Loss_{P}(x)$ indicates the total loss
of $P$ on $x$, i.e. the sum of
all $n$ losses:
$\Loss_{P}(x) = \sum_{i=0}^{|x|-1}(-\log|P(x_1\dots x_i)-1+x_{i+1}|)$.
Thus, the snooping curve $L_x(\alpha)$ gives 
the minimal loss suffered on all of $x$ by
a prediction strategy, as a function of the complexity at most $\alpha$
of the contemplated class of prediction strategies. 
The question arises what shapes these functions can have---for example,
whether there can be sharp drops in the loss for only minute
increases in complexity of prediction strategies.

A result of~\cite{Vy01} describes possible shapes of $L_x$ but only
for $\alpha=o(n)$ where $n$ is the length of $x$. 
    Here we show that for every function $L$ and every $k\le n$
    there is a data sequence $x$ such that
    $L_x (\alpha \pm O(\log n)) = L( \alpha)  \pm O(\log n)$, provided
    $L(0)\le n$, $L(\alpha)+\alpha$ is non-increasing on $[0,k]$, and 
    $L(\alpha)=0$ for  $\alpha\ge k$.
    \begin{lemma}
    $L_x(\alpha \pm O(\log n))=h_x(\alpha \pm O(\log n))$
    for every $x$ and $\alpha$. Thus, Lemma~\ref{lem.equiv} and Theorem~\ref{th14}
    describes also the coarse shape of all possible snooping curves.
    \end{lemma}
\begin{proof}
($\leq$) A given finite set $A$
of binary strings of length $n$ can be identified
with the following prediction strategy $P$:
Having read the prefix $y$ of $x$ it outputs $p=|A_{y1}|/|A_y|$ where
$A_y$ stands for the number of strings in $A$ having prefix $y$.

It is easily seen, by induction,
that $\Loss_P(y)=\log(|A|/|A_y|)$ for every $y$. Therefore,
$\Loss_P(x)=\log|A|$ for every $x\in A$. Since $P$
corresponds to $A$ in the sense that $K(P \mid A)= O(1)$,
we obtain $L_x(\alpha+O(\log n))\le h_x(\alpha)$.
The term $O(\log n)$ is required, because
the initial set of complexity $\alpha$ might
contain strings of different lengths while we need to know $n$ to get rid of 
the strings of lengths different from $n$.   

($\geq$) Conversely, assume that 
$\Loss_P(x)\le m$. Let $A=\{x\in\{0,1\}^n \colon\Loss_P(x)\le m\}$.
Since $\sum_{|x|=n} 2^{-\Loss_P(x)} = 1$ 
(proof by induction on $n$), and $2^{-\Loss_P(x)}\ge2^{-m}$
for every $x\in A$, we can conclude that $A$ has at most $2^{m}$ elements.   
Since $K(A \mid P)=O(\log m)$, we obtain $h_x(\alpha+O(\log n))\le L_x(\alpha)$.
\end{proof}
Thus, within the obvious constraint of the function $L_x (\alpha)+ \alpha$
being non-increasing, all shapes for the minimal total loss $L_x (\alpha)$ as
a function of the allowed predictor complexity  are possible.

\subsection{Foundations of MDL}
\label{ex.MDL}
   (i) Consider the following algorithm based on the Minimum Description
   Length principle. Given $x$, the data to explain, and $\alpha$,
   the maximum allowed complexity of explanation, we search for programs $p$
   of length at most $\alpha$ that print a finite set $S\ni x$. Such
   pairs $(p,S)$ are possible explanations. 
   The {\em best explanation} is defined to be
   the $(p,S)$ for
   which $\delta(x|S)$ is minimal. Since the function
   $\delta(x|S)$ is not computable, we cannot find
   the best explanation.
   The programs use unknown
   computation time and thus we can never be certain that we have
   found all possible explanations.

   To overcome this problem we use the indirect method of MDL: 
   We run all
   programs in dovetailed fashion. 
   At every computation step $t$ consider all pairs $(p,S)$ such that
   program $p$ has printed the set $S$ containing $x$ by time $t$.
   Let $(p_t,L_t)$ stand for 
   the pair
   $(p,S)$ such that $|p|+\log|S|$ is minimal among all these pairs $(p,S)$.
   The best hypothesis $L_t$ changes from time to time
   due to the appearance of a better hypothesis. Since no hypothesis is
   declared best twice, from some moment onwards the explanation
   $(p_t,L_t)$ which is declared best does not change anymore.

   Compare this indirect method with the direct one:
   after step $t$ of dovetailing
   select
   $(p,S)$ for which $\log|S|-K^t(x|S)$ is minimum among all programs $p$ that
   up to this time have printed a set  $S$ containing $x$,
   where $K^t(x|S)$ is the approximation of  $K^t(x|S)$
   obtained after $t$ steps of dovetailing, that is,
   $K^t(x|S)=\min\{|q|: U \text{ on input }\pair{q,S}\text{ prints } x
   \text{ in at most } t\text{ steps}\}$.
   Let $(q_t,B_t)$ stand for that model.
   This time the same hypothesis
   can be
   declared best twice. However from some moment onwards the explanation
   $(q_t,B_t)$ which is declared best does not change anymore.

   Why do we prefer the indirect method to the direct one?
   The explanation is that we have a comparable situation in the
   practice of the real-world MDL, in the analogous
   process of finding the MDL code. There,
   we deal often with $t$ that are much less than the time of
   stabilization of both $L_t$ and $B_t$. For small $t$, the model $L_t$
   is better than $B_t$ in the following
   respect: $L_t$ has some guarantee of goodness, as we know that
   $\delta(x|L_t) + K(x) \le |p_t|+\log|L_t| + O(1)$. That is, we know
   that the sum of deficiency of $x$ in $L_t$ and $K(x)$ is less than some
   known value. In contrast, the model $B_t$
   has no guarantee of goodness at all:
   we do not know any upper bound neither for $\delta(x|B_t)$, nor for
   $\delta(x|B_t) + K(x)$.

   Theorem~\ref{th13} implies that the
   indirect method of MDL gives not only some garantee of goodness
   but also that, in the limit, that guarantee approaches the value
   it upper bounds, that is, approaches $\delta(x|L_t) + K(x)$, and
   $\delta(x|L_t)$ itself is not much greater than $\delta(x|B_t)$
   (assuming that $\alpha$ is not critical). That is, in the limit,
   the method of MDL will yield an explanation that is only a little
   worse than the best explanation.

(ii) If $S \ni x$ is a smallest set such that  $K(S) \leq \alpha$, then
$S$ can be converted into a best strategy of complexity at most $\alpha$,
to predict the successive bits of $x$
given the preceding ones, (Section~\ref{ex.snoop}).
Interpreting ``to explain'' as ``to be able to predict well'',
MDL in the sense of sets witnessing $\lambda_x (\alpha)$ 
gives indeed a good explanations at every complexity level $\alpha$.


    (iii) In statistical applications of MDL \cite{Ri83,BRY},
    MML \cite{Wa87}, and related methods, one selects the model in a given 
    model class that
    minimizes the sum of the model code length and the data-to-model
    code length; in modern versions \cite{BRY} one 
    selects the model that 
    minimizes just the data-to-model code length 
    (ignoring the model code length). For example, one uses data-to-model code 
    $- \log P(x)$ for data $x$ with respect to probability (density function)
    model $P$. For example, if the model is the uniform distribution
    over $n$-bit strings, then the data-to-model code for $x=00 \ldots 0$
    is $- \log 1/2^n = n$, even though we can compress $x$ to about
    $log n$ bits, without even using the model. Thus, the data-to-model code
    is the worst-case number of bits required for data of given length using
    the model, rather than the optimal number of bits for the 
    particular data at hand. This is precisely what we do in the
    structure function approach: the data-to-model cost of $x$ with respect 
    to model $A \ni x$ is $\log |A|$, the worst-case number of bits required
    to specify an element of $A$ rather than the minimal number of bits
    required to specify $x$ in particular.
    In contrast,
    ultimate compression of the two-part code, which is suggested 
    by the ``minimum description length'' phrase, \cite{ViLi99},
    means minimizing
    $K(A)+K(x|A)$ over all models $A$ in the model class.
    In Theorem~\ref{th13} we have essentially shown that the ``worst-case''
    data-to-model code above is the approach that guarantees the best
    fitting model. In contrast, the ``ultimate compression'' approach
    can yield models that are far from best fit. (It is easy to see
    that this happens only
    if the data are ``not typical'' for the contemplated model, \cite{ViLi99}.) 
    For instance, let $x$ be a string of length $n$ and complexity about $n/2$
    for which $\beta_x(O(\log(n))=n/4+O(\log(n)$. This means that
    the best model at a very low complexity level (essentially level 0
    within the ``logarithmic additive precision'' which governs our
    techniques and results) has significant randomness deficiency and
    hence is far from ``optimal'' or ``sufficient''.  Such strings
    exist by Corollary~\ref{cor.beta}. Such strings are not
    the strings of maximal Kolmogorov complexity, with $K(x) \geq n$,
    such as most likely result from $n$ flips with a fair coin, but
    strings that must have a more complex cause since their minimal
    sufficient statistic has complexity higher than $O(\log n)$.
    Consider the model class
    consisting of the finite sets containing $x$ at complexity
    level $\alpha=O(\log n)$. Then for the model $A_0=\{0,1\}^n$ we have
    $K(A_0)=O(\log n)$ and $K(x|A_0)=n/2+O(\log n)$ thus
    the sum $K(A_0)+K(x|A_0)=n/2+O(\log n)$ is minimal up to
    a term $O(\log n)$.
    However, the randomness defficiency of $x$ in $A_0$ is
    about $n/2$, which is much bigger than the minimum 
    $\beta_x(O(\log(n))\approx n/4$.
    For the model $A_1$ witnessing
    $\beta_x(O(\log(n))\approx n/4$ we also have
    $K(A_1)=O(\log n)$ and $K(x|A_1)=n/2+O(\log n)$. 
    However, it has smaller cardinality:
    $\log|A_1|=3n/4+O(\log n)$ which causes the smaller randomness
    deficiency.

    The same happens also for other model classes, such as
    probability models, see Appendix~\ref{app.extension}.
    Consider, for instance, 
    the class of Bernoulli
    processes with rational bias $p$ for outcome ``1''
    ($0 \leq p \leq 1$) to generate binary strings of length $n$.
    Suppose
    we look for the model minimizing the codelength
    of the model plus data given the model: $K(p|n)+K(x|p,n)$.
    Let the data be $x= 00 \ldots 0$.
    Then the probability model $P$ (the uniform distribution)
    with $P(x)=1/2^n$ corresponding to
    probability $p = \frac{1}{2}$ compresses the data code
    to $K(x \mid n,p) = O(1)$ bits since we can describe $x$
    by the program {\tt print n ``0''s}, and hence need only $O(1)$
    bits apart from $n$.  We also trivially have $K(p|n) \leq K(p)+O(1) =O(1)$. 
    But we cannot distinguish between the probability model $P$ 
    hypothesis based on $p$
    and the probability model $P'$ with $P'(x)=1$ (singular distribution)
    hypothesis based on $p'$ in terms of tthese code lengths:
    we find the same code length 
    $K(x \mid n,p') = O(1)$ bits and $K(p'|n)=O(1)$
    if we replace $p= \frac{1}{2}$ by $p'=0$ in these expressions.
    Thus we have no basis to prefer hypothesis
    $p$ or hypothesis $p'$, even though
    the second possibility is overwhelmingly more likely.
    This shows that ultimate compression of the two-part code,
    here for example resulting in $K(p|n)+K(x|n,p)$,
    may yield a (probability) model $P$ based on $p=\frac{1}{2}$ for which 
    the data has the maximal possible randomness deficiency
    ($- \log P(x) - K(x \mid n,p) = n-O(1)$ and hence
    is atypical.

    However, in the structure functions $h_x(\alpha)$ and $\lambda_x(\alpha)$
    the data-to-model code  for the
    model $p = \frac{1}{2}$ is
    $- \log P(x) = - \log (\frac{1}{2})^n = n$ bits,
    while $p' = 0$
    results $- \log P'(x)= - \log 1^n = 0$ bits. 
    Choosing the shortest data-to-model code
    results in the minimal randomness deficiency, as in
    (the generalization to probability distributions of) Theorem~\ref{th13}.

(iv) Another question arising in MDL or maximum likelihood (ML) estimation
is its performance if the ``true'' model is
not part of the contemplated  model class. Given certain data, why would 
we assume they are generated by probabilistic or deterministic processes?
They have arisen by natural processes most likely 
not conforming to mathematical
idealization. Even if we can assume the data arose from a process
that can be mathematically formulated,
such situations arise if
we restrict modeling of data arising from a ``complex'' source
(conventional analogue being data arising from $2k$-parameter sources) by
``simple'' models (conventional analogue being $k$-parameter models).
Again, Theorem~\ref{th13} shows that, within the class of models
of maximal complexity $\alpha$, these constraints
we still select a simple model for which the data is maximally typical.
This is particularly significant for data $x$ if the allowed complexity $\alpha$
is significantly
below the complexity of the Kolmogorov minimal sufficient statistic,
that is, if $h_x (\alpha)+\alpha \gg K(x)+c$.
This situation is potentially common, for example if we have
a small data sample generated by a complex process. Then,
the data will typically be non-stochastic in the sense
of  Section~\ref{ex.nonstoch}. For a data sample that is very large 
relative to the complexity of the process generating it, this will
typically not be the case and the structure function will drop
to the sufficiency line early on.

\subsection{Foundations of Maximum Likelihood}
            \label{ex.ML}
The algorithm based on ML principle
is similar to the algorithm of the previous example.
The only difference is that the  currently best
$(p,S)$ is the one for which $\log|S|$ is minimal.
In this case the limit hypothesis
$\tilde S$ will witness $h_x(\alpha)$ and we obtain the same
corollary:
$\delta(x|S)\le\beta_x(\alpha-O(\log n))+O(\log n)$.

 \subsection{Approximation Improves Models}
   Assume that in the MDL algorithm, as
   described in Section~\ref{ex.MDL}, we change the currently best explanation
   $(p_1,S_1)$ to the explanation
   $(p_2,S_2)$ only if $|p_2|+\log|S_2|$
   is much less than  $|p_1|+\log|S_1|$,
   say $|p_2|+\log|S_2|\le|p_1|+\log|S_1|-c\log n$ for a constant $c$.
   It turns out that if $c$ is large enough and $p_1$ is
   a shortest program of $S_1$, then
   $\delta (x \mid S_2)$ is much less than  $\delta (x \mid S_1)$.
   That is, every time we change the explanation we improve
   its goodness unless the change is just caused by the fact that
   we have not yet found the minimum length program for the current model.
   \begin{lemma} 
   There is a constant $c$ such that
   if
   $\Lambda (S_2) \le \Lambda (S_1) -2c \log |x|$,
   then
   $\delta (x \mid S_2) \le \delta (x \mid S_2)-c\log|x|+O(1)$.
   \end{lemma}

   \begin{proof}
   Assume the notation of Theorem~\ref{th13}.
   By (\ref{eq84}), for every pair of sets $S_1,S_2 \ni x$ we have
   $\delta (x \mid S_2 ) - \delta (x \mid S_1) =
   \Lambda (S_2) - \Lambda (S_1) +  \Delta$
   with $\Delta = K(S_1 \mid x^*)  - K(S_2 \mid x^*) + O(1) \leq
   K(S_1 \mid S_2,x^*)+O(1)\le K(S_1 \mid S_2,x)+O(1)$.
   As $\Lambda (S_2) - \Lambda (S_1)\le
   |p_2|+\log|S_2|- \Lambda (S_1)=|p_2|+\log|S_2|-(|p_1|+\log|S_1|)\le
   -2c\log|x|$ we need to prove that
   $K(S_2 \mid S_1,x)\le c\log|x|+O(1)$.
   Note that $(p_1,S_1)$, $(p_2,S_2)$ are consecutive explanations
   in the algorithm and every explanation may appear only once.
   Hence to identify $S_1$ we only need to know
   $p_2,S_2,\alpha$ and $x$. Since $p_2$ may be found from $S_2$ and
   length $|p_2|$ as the first program computing $S_2$ 
   of length $|p_2|$, obtained by running
   all programs dovetailed style, we have
   $K(S_2 \mid S_1,x)\le 2\log|p_2|+2\log|\alpha|+O(1)\le 4\log |x|+O(1)$.
   Hence we can choose $c=4$.
   (Continued in Section~\ref{ex.prnr}.)
   \end{proof}

   \subsection{Non-stochastic Objects} 
   \label{ex.nonstoch}
   Let $\alpha_0,\beta_0$ be natural numbers.
   A string $x$ is called  
   {\em $(\alpha_0,\beta_0)$-stochastic} by Kolmogorov
   if $\beta_x(\alpha_0)\le\beta_0$.
   In \cite{Sh83} it is proven that for some
   $c,C$ for all $n$ and all $\alpha_0,\beta_0$ with
   $2\alpha_0+\beta_0< n-c\log n-C$
   there is a string $x$ of length
   $n$ that is not  $(\alpha_0,\beta_0)$-stochastic.
   Corollary~\ref{cor.beta} strengthens this result of Shen:
   for some
   $c,C$ for all $n$ and all $\alpha_0,\beta_0$ with
   $\alpha_0+\beta_0< n-c\log n-C$
   there is a string $x$ of length
   $n$ that is not  $(\alpha_0,\beta_0)$-stochastic.
   Indeed, apply Corollary~\ref{cor.beta}
   to $k=\alpha_0+c_1\log n+C_1$ (we will choose $c_1,C_1$ later)
   and the function $\beta(i)=n-k$ for $i<k$ and
   $\beta(i)=0$ for $i=k$.
   For the $x$ existing by Corollary~\ref{cor.beta} we have
   $\beta_x(\alpha_0)\ge 
   \beta(\alpha_0\pm(c_2\log n+C_2))-(c_2\log n+C_2)
   \ge\beta(k-1)-(c_2\log n+C_2)=
   n-k-(c_2\log n+C_2)=n-(\alpha_0+c_1\log n+C_1)-(c_2\log n+C_2)>\beta_0$.
   (The first inequality is true if $\alpha_0+c_2\log n+C_2\le k-1$; thus
   let $c_1=c_2,C_1=C_2+1$.
   For the last inequality to be true let
   $c=c_1+c_2$ and $C=C_1+C_2$.)
   That is, $x$ is not $(\alpha_0,\beta_0)$-stochastic.

\section{Fine Structure and Sufficient Statistic}
\label{sect.fine}

Above, we looked at the coarse shape of the structure function,
but not at the fine detail. 
We show that $h_x$ coming from infinity drops 
to the sufficiency line $L$ defined by 
   $L (\alpha)+\alpha = K(x)$. It first touches this line for
   some $\alpha_0\le K(x)+O(1)$.
It then touches this
line a number of times (bounded by a universal constant) and in between
moves slightly (logarithmically) away in little bumps.
There is a simple explanation why these bumps are there:
   It follows from (\ref{eq.soi}) and \eqref{eq57}
   that there is a constant $c_1$
   such that for every $S \ni x$, we have
   $K(S)+\log|S| \ge K(x)+K(S \mid x^*) - c_1$.
   If, moreover,
   $K(S)+\log|S| \le
   K(x)+c_2$, then
   $K(S \mid x^*) \le c_2+c_1$. This
   was already observed in \cite{GTV01}.
   Consequently,
   there are less than $2^{c_2+c_1+1}$ distinct such
   sets $S$. Suppose
   the graph of $h_x$ drops
   within distance $c_2$ of the sufficiency line at $\alpha_0$, then
   it cannot be within distance $c_2$ on more than $2^{c_2+c_1+1}$ points.
   By the pigeon-hole principle,
   there is $\alpha\in[\alpha_0,K(x)]$
   such that $h_x(\alpha)+\alpha\ge\lambda_x(\alpha)\ge
   K(x)+\log (K(x)-\alpha_0)-c_2-1$.
   So if $|K(x)-\alpha_0|$ is
   of order $\Omega(n)$ , then
   we obtain the logarithmic bumps, or possibly only one logarithmic
   bump, on the interval $[\alpha_0,K(x)]$.
   However, we  will show below that $h_x$ cannot move away
   more than $O(\log |K(x)-\alpha_0|)$ from
   the sufficiency line on the interval $[\alpha_0,K(x)]$.
   The intuition here is that a data sequence can have a simple
satisfactory probabilistic explanation, but we can also explain
it by many only slightly more complex explanations that are slightly
less satisfactory but also model more
accidental random features---models that are only slightly more
complex but that significantly overfit the data sequence by
modeling noise.

\subsection{Initial behavior}
Let $x$ be a string of complexity $K(x)=k$.
The structure function $h_x (\alpha)$ defined by (\ref{eq2})
rises sharply above the 
sufficiency line for  
very small values of $\alpha$ with 
$h_x (\alpha) = \infty$ for $\alpha$ close to 0.
   To analyze the behavior of $h_x$ near the origin, define
   a function
   \begin{equation}\label{eq.bb-1}
   m (x) = \min_y \{K(y): y \geq x \},
   \end{equation}
   the minimum complexity of a string greater than $x$---that is,
   $m(x)$ is the greatest monotonic non-decreasing function
   that lower bounds $K(x)$. The function $m(x)$ tends to infinity
   as $x$ tends to infinity, very  slowly---slower than any
   computable function.

   For every $\alpha \in [0, m(x)-O(1))$ we have $h_x (\alpha) = \infty$.
   To see this, we reason as follows:
   For a set $S \ni x$ with $K(S)= \alpha$ with $\alpha$ in the above range
   we can consider
   the largest element $y$ of $S$.
   Then $y$ has complexity $\alpha +O(1)<m(x)$,
   that is, $K(y) < m(x)$, which implies that $y< x$. But
   then $x \not\in S$ which is a contradiction.

\subsection{Sufficient Statistic} 
A sufficient statistic of the data
contains all information in the data about the model.
In introducing the notion of sufficiency in classical
statistics, Fisher~\cite{Fi22} stated:
       ``The statistic chosen should summarize the whole of the relevant
information supplied by the sample. This may be called
       the Criterion of Sufficiency $\ldots$
In the case of the normal curve
of distribution it is evident that the second moment is a
       sufficient statistic for estimating the standard deviation.''
For the classical (probabilistic) theory see, for example, \cite{CT91}. In
\cite{GTV01} an algorithmic theory of sufficient
statistic  (relating individual data
to individual model) was developed
and its relation with the probabilistic version established.
The algorithmic basics are as follows:
 Intuitively, a model expresses the essence
of the data if the two-part code describing the data consisting of the
model and the data-to-model code is as concise as the best one-part description.
Formally:
\begin{definition}
\rm
A finite set  $S$ containing $x$ is \emph{optimal for $x$} if  
                     \begin{equation}\label{eq.ss}
\Lambda(S)\le K(x)+c.
                     \end{equation}
Here $c$ is some small value, constant or logarithmic in $K(x)$,
depending on the context. A minimal length description $S^*$ of such an
optimal set is called a \emph{sufficient statistic for $x$}. To
specify the value of $c$ we will say \emph{$c$-optimal} and
\emph{$c$-sufficient}. 
\end{definition}

If a set $S$ is $c$-optimal with $c$ constant, then by
(\ref{eq.descr}) we have $K(x) - c_2 \leq \Lambda(S) \leq K(x)
+c$. Hence, with respect to the structure function
$\lambda_x(\alpha)$ we can state that all optimal sets $S$ and
only those, cause the function $\lambda_x$ to drop to its minimal
possible value $K(x)$. We know that this happens for at least one
set, $\{x\}$ of complexity $K(x)+O(1)$.
 
We are interested in finding  optimal  sets  that  have  low
complexity.  Those   having   minimal	complexity   are   called
\emph{minimal optimal sets}  (and  their  programs  \emph{minimal
sufficient statistics}). The less optimal the sets
are, the more additional noise in the data they start to model,
see the discussion of overfitting in the initial paragraphs of
Section~\ref{sect.coarse}. To be rigorous  we  should  say  minimal
among $c$-optimal. We know from \cite{GTV01} that the complexity
of a minimal optimal set is at least $K(K(x))$, up to a fixed
additive constant, for every $x$.
So for smaller
arguments the structure function definitively rises above the
sufficiency line. 
   We also know
   that for every $n$ there are so-called {\em non-stochastic}
   objects $x$ of length $n$ that have optimal sets
   of high complexity only. For example, there are
   $x$ of complexity $K(x \mid n^* ) = n +O(1)$
   such that every optimal set $S$
   has also complexity $K(S \mid n^*) = n + O(1)$, hence by
   the conditional version $K(S \mid n^*)+ \log |S| \le K(x \mid n^*)+c$
   of (\ref{eq.ss}) we find $|S|$ is bounded by a fixed universal constant.
   As $K(S \mid x^*)=O(1)$ (this is proven in the beginning of this section),
   for every $y \in S$ we have
   $K(y \mid x^*) \le K(y \mid S)+K(S \mid x^*)+ O(1)=O(1)$.
   Roughly speaking for such $x$ there is no other optimal set $S$ than
   the singleton $\{x\}$.
\begin{example}
\rm
{\bf Bernoulli Process:}
Let us look at the coin toss example 
of Item (iii) in Section~\ref{ex.MDL}, this time in the sense
of finite set models rather than probability models.
Let $k$ be a number in the range $0,1,\dots,n$ 
of complexity $\log n+ O(1)$ given $n$ and let $x$ be a string of length 
$n$ having $k$ ones of complexity $K(x \mid n,k) \geq \log {n \choose k}$ 
given $n,k$. This $x$ can be viewed as a typical result of 
tossing a coin with a bias about $p=k/n$.
A two-part description
of $x$ is given by
the number $k$ of 1's in $x$ first, followed by the index
$j \leq \log |S|$  of $x$ 
in the set $S$ of strings of length $n$ with $k$ 1's. 
This set is optimal, since 
$K(x \mid n)=K(x,k \mid n)=K(k \mid n)+K(x \mid k,n)= K(S|n)+ \log|S|$.
\end{example}

   \begin{example}\label{ex.structure}
   \rm
   {\bf Hierarchy of Sufficient Statistics:}
   Another possible application of the theory is to find a good
   summarization of the meaningful information
   in a given picture.
   All the
   information in the picture is described by a binary string $x$
   of length $n=ml$ as follows.
   Chop $x$ into $l$
   substrings $x_i$ ($1 \leq i \leq l$)
   of equal length $m$ each. Let $k_i$ denote the number of
   ones in $x_i$. Each such substring metaphorically
   represents a patch of, say, color.
   The intended color, say ``cobalt blue'', is indicated by the number of ones in
   the substring. The actual color depicted may be typical cobalt blue
   or less typical cobalt blue.
   The smaller the randomness deficiency
   of substring $x_i$ in the set
   of all strings of length $m$ containing precisely $k_i$ ones,
   the more typical $x_i$ is, the better it achieves a typical
   cobalt blue color.
   The metaphorical ``image'' depicted by $x$
   is $\pi (x)$, defined as the string $k_1k_2\dots k_l$ over the
   alphabet $\{0,1,\dots,m\}$,
   the set of colors available.
   We can now consider several statistics for $x$.

   Let $X \subseteq\{0,1,\dots,m\}^l$  (the set of possible
   realizations of the target image),
   and let $Y_i$ for $i=0,1,\dots,m$ be
   a set of
   binary strings of length $m$ with $i$ ones (the set
   of realizations of
   target color $i$). Consider the set

           $$
   S=\{x' : \pi(x')\in X,
   (x')_i\in Y_{k_i} \text{ for all }i=1,\dots,l\}
           $$
   One possible application of these ideas are
   to gouge how good the picture is with
   respect to the given summarizing set $S$.
   Assume that $x\in S$. The set  $S$ is then a
   statistic for $x$ that captures both the colors of the patches
   and the image, that is, the total picture. If
   $S$ is a sufficient statistic of $x$
   then $S$ perfectly expresses the meaning aimed for by
   the image and the true color aimed for in everyone of the color patches.
   Clearly,
   $S$ summarizes the relevant
   information in $x$ since it captures both image and coloring, that is,
   the total picture.  But we
   can distinguish more sufficient statistics.

   The set
           $$
   S_1=\{x' : \pi(x')\in X\}
           $$
   is a statistic that captures only the image.
   It can be sufficient only if all colors used in the picture $x$ are
   typical.
   The set
           $$
   S_2 =\{x': (x')_i\in Y_{k_i} \text{ for all }i=1,\dots,l\}
           $$
   is a statistic that captures the color information in the picture.
   It can be sufficient only if the image is a random string of length
   $l$ over the alphabet $\{0,1,\dots,m\}$, which is surely not the case for
   all the real images.
   Finally the set
           $$
   A_i =\{x' : (x')_i\in Y_{k_i}\}
           $$
   is a statistic that captures only the color of patch $(x')_i$
   in the picture.
   It can be sufficient only if $K(i)\approx0$ and
   all the other color applications and the image are typical.
   \end{example}

\subsection{Bumps in the Structure Function}
\label{ex.bumps}
Consider $x \in \{0,1\}^n$ with $K(x \mid n)= n+ O(1)$ and  the
conditional variant 
$h_x (\alpha \mid y ) = \min_S \{ \log |S| : S \ni x, \; |S| < \infty ,
\; K(S \mid y) \leq \alpha \}$ of (\ref{eq2}).
Since $S_1=\{0,1\}^n$ is a set containing $x$  and
can be described by $O(1)$ bits (given $n$),  
we find $h_x (\alpha  \mid n)  \leq n+O(1)$ for $\alpha = K(S_1\mid n)=O(1)$. 
For increasing $\alpha$,
the size of a set $S \ni x$,  one can describe in 
$\alpha$ bits, decreases monotonically until
for some $\alpha_0$ we obtain a first set $S_0$
witnessing $h_x(\alpha_0 \mid n)+ \alpha_0 = K(x\mid n) + O(1)$. 
Then, $S_0$ is a minimal-complexity optimal set for $x$,
and $S^*_0$ is a minimal sufficient statistic for $x$.
Further increase of $\alpha$ halves the
set $S$ for each additional bit of $\alpha$ until $\alpha =K(x\mid n)$. 
In other words, for every increment $d$ we have 
$h_x(\alpha_0+d \mid n) = K(x\mid n) - (\alpha_0 +d + O(\log d)) $,
provided the right-hand side is non-negative, and 0 otherwise.
Namely, once we have an optimal set $S_0$ we can subdivide it in a
standard way into $2^{d}$ parts and take as new set $S$ the part
containing $x$. The $O(\log d)$ term is due to the fact that
we have to consider self-delimiting encodings of $d$. 
This additive term  is there to stay, it cannot be eliminated.
For
$\alpha  \geq K(x\mid n)$ obviously the smallest set $S$ containing $x$
that one can describe 
using $\alpha$ bits (given $n$) is the singleton set $S= \{x\}$.
The same analysis can be given for the unconditional version
$h_x(\alpha)$ of the structure function, which behaves the same except
for possibly the small initial part $\alpha \in [0,K(n))$ where
the complexity is too small to specify the set $S_1=\{0,1\}^n$,
see the initial part of Section~\ref{sect.fine}.

\begin{figure}
\begin{center}
\epsfxsize=8cm
\epsfxsize=8cm \epsfbox{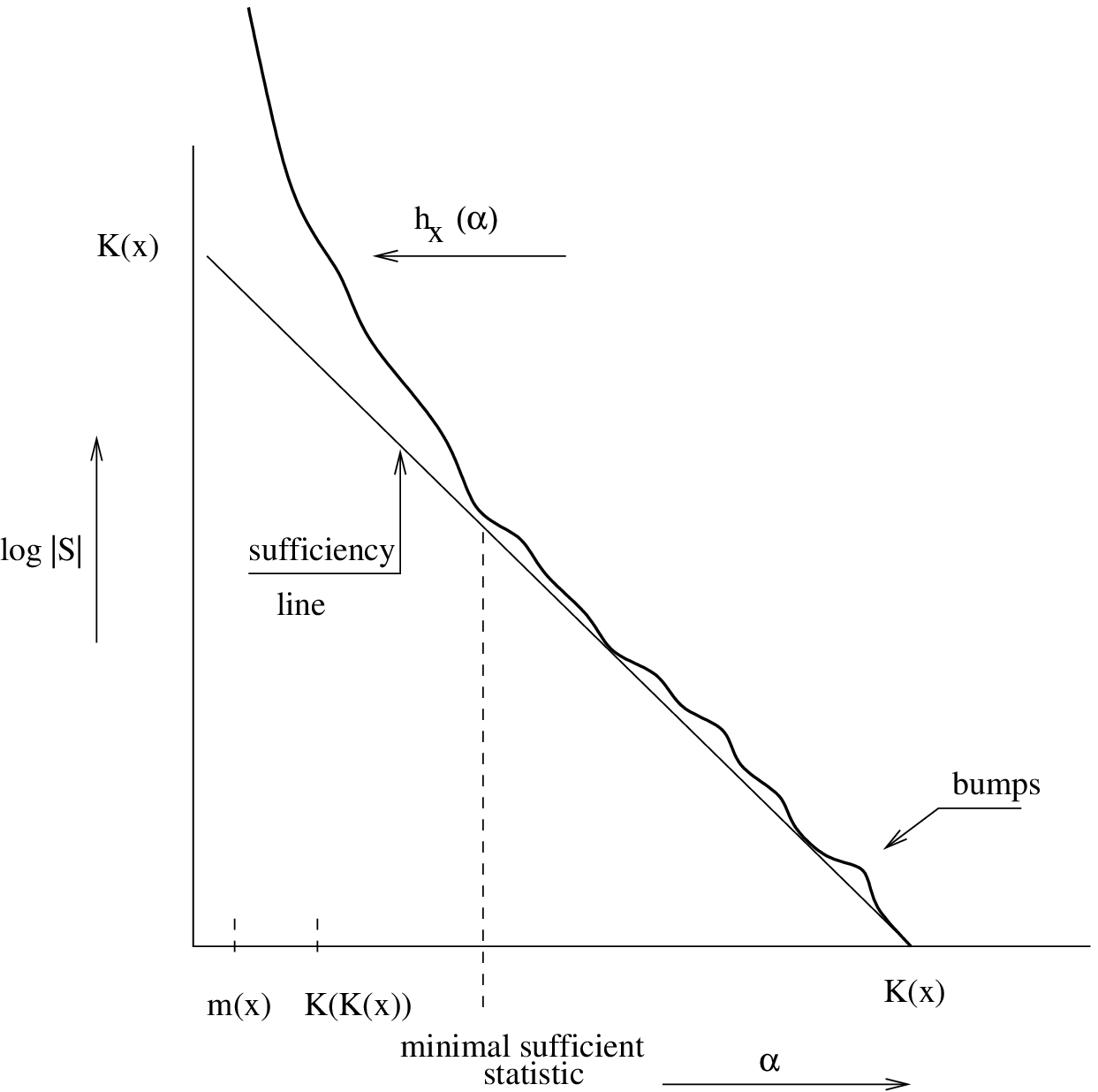}
\end{center}
\caption{Kolmogorov structure function}
\label{fig.ksf}
\end{figure}

The little bumps in the sufficient statistic region 
$[K(K(x)), K(x)]$ in Figure~\ref{fig.ksf} are due
to the boundedness of the number of sufficient statistics.

\subsection{``Positive'' and ``Negative'' Randomness}
\label{ex.prnr}
\begin{figure}
\begin{center}
\epsfxsize=8cm
\epsfxsize=8cm \epsfbox{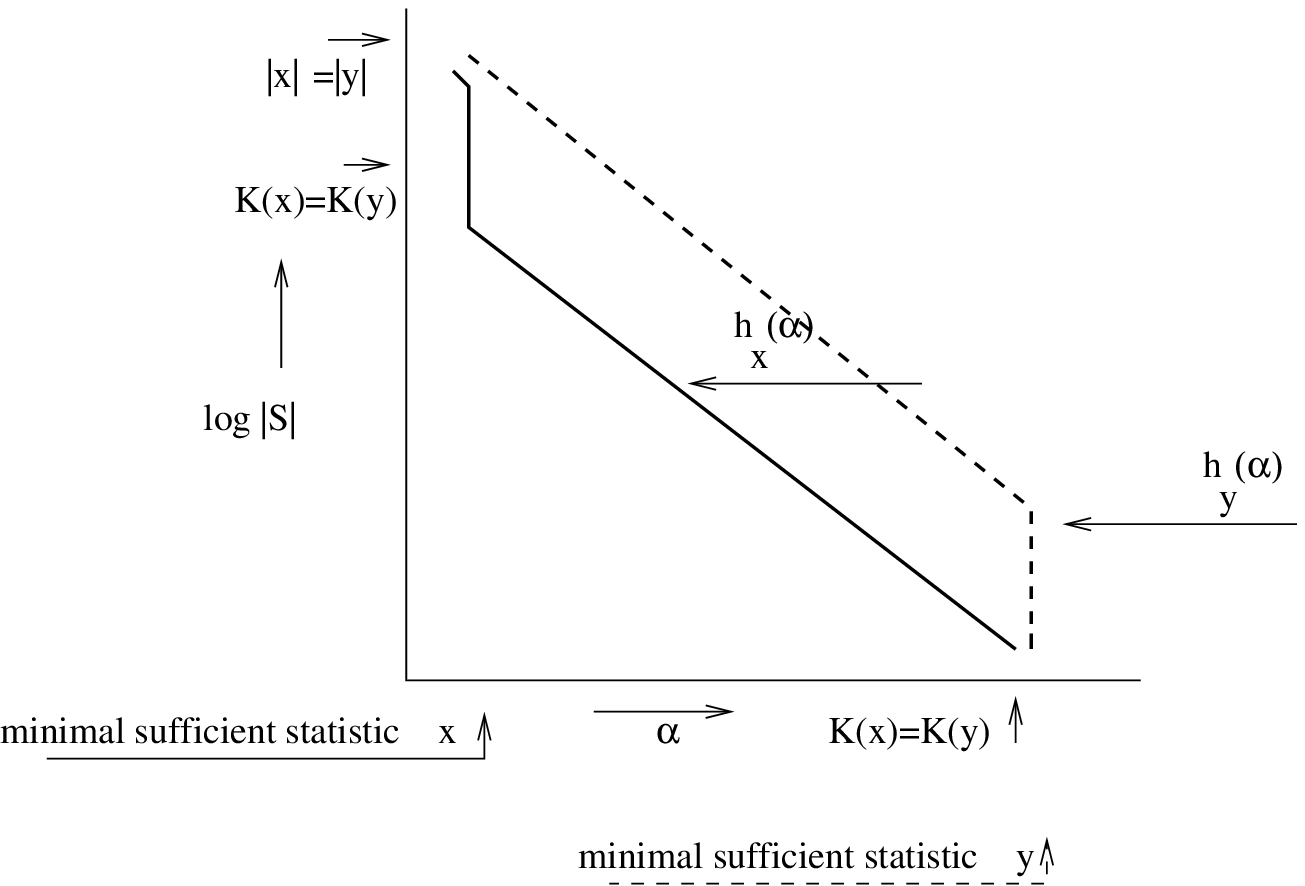}
\end{center}
\caption{Data string $x$ is ``positive random'' or ``stochastic''
 and data string $y$
is only ``negative random'' and ``non-stochastic''.}
\label{figure.pos_negrandom}
\end{figure}

(Continuing Section~\ref{ex.nonstoch}.)
   In \cite{GTV01} the existence
   of strings was shown for which essentially
   the singleton set consisting of the string itself is a minimal
   sufficient statistic. While a sufficient
statistic of an object yields a two-part code that is as short as the shortest
one part code, restricting the complexity of the allowed statistic
may yield two-part codes that are considerably longer than the best one-part 
code (so the statistic is insufficient). 
This is what happens for the non-stochastic objects. In fact,
for every object there is a complexity bound below which this happens---but
if that bound is small (logarithmic) we call the object ``stochastic''
since it has a simple satisfactory explanation (sufficient statistic). 
Thus, Kolmogorov in \cite{Ko74b} (full text given in Section~\ref{sect.trans})
 makes the important distinction of
an object just being random in the ``negative'' sense by having 
high Kolmogorov complexity,
and an object having high Kolmogorov complexity 
but also being random in the ``positive,
probabilistic'' sense of having a low-complexity minimal sufficient statistic.
An example of the latter is a string $x$ of length $n$ with $K(x) \geq n$, 
being typical for the
set $\{0,1\}^n$, or the uniform probability distribution over that set,
while this set or probability distribution 
has complexity $K(n)+O(1) = O(\log n)$.
We depict the distinction in Figure~\ref{figure.pos_negrandom}.

   Corollary~\ref{cor.beta} establishes that for some constant $C$, for
   every length $n$, for every
   complexity $k \leq n$
   and every $\alpha_0 \in [0,k]$,
   there are $x$'s of length $n$ and complexity $k\pm C\log n$ such that
   the minimal randomness deficiency $\beta_x (i) >  n-k-C\log n$
   for every $i \leq \alpha_0 - C\log n$ and $\beta_x (i) < C\log n$
   for every $i \geq \alpha_0 +C\log n$.
   Fix $\eps=C\log n$ and
   define for all $s,t=0,\dots,n/(2\eps)-1$ the
   set $A_{st}$ of all
   $n$-length strings
   of complexity
   $K(x)\in[(2s-1)\eps,(2s+1)\eps)$ and
   such that
   the minimal randomness deficiency $\beta_x (i) >  n-(2s+1)\eps$
   for every $i \leq (2t-1)\eps $ and $\beta_x (i) <\eps$
   for every $i \geq (2t+1)\eps $.
   Corollary~\ref{cor.beta} implies that
   every $A_{st}$ is  non-empty (let $\alpha_0=2t\eps$, $k=2s\eps$).
   Note that
   $A_{st}$ are pair-wise disjoint. Indeed, if $s\ne s'$ then
   $A_{st}$ and $A_{s't'}$ are disjoint as the corresponding
   strings $x,x'$ have different complexities.
   And if $t\ne t'$, say $t<t'$, then
   $A_{st}$ and $A_{s't'}$ are disjoint, as the corresponding   strings
   $x,x'$ have different value of deficiency function
   in the point $i=(2t+1)\eps$:
   $\beta_x ((2t+1)\eps) >  n-(2s+1)\eps\ge \eps>\beta_{x'}((2t+1)\eps)$.

   Letting $k=\alpha_0=n-\sqrt n$ we see
   that there are $n$-length non-stochastic
   strings of almost maximal complexity $n -  \sqrt{n}\pm O(\log n)$
   having significant $\sqrt{n}\pm O(\log n)$ randomness deficiency with
   respect to $\{0,1\}^n$ or, in fact, every other finite set
   of complexity less than $n - O(\log n)$!

\section{Computability Questions}
\label{sect.comput}

How difficult is it to compute the functions $h_x, \lambda_x, \beta_x$,
and the minimal sufficient statistic? To express the properties
appropriately we require the notion of functions
that are not computable,
but can be approximated monotonically by a computable
function. 
\begin{definition}
\rm
\label{def.semi}
A function $f: {\cal N} \rightarrow {\cal R}$ is
{\em upper semi-computable} if there is a Turing machine $T$ computing a
total function $\phi$
such that $\phi (x,t+1) \leq \phi (x,t)$ and
$\lim_{t \rightarrow \infty} \phi (x,t)=f(x)$. This means
that $f$ can be computably approximated from above.
If $-f$ is upper semi-computable, then $f$ is lower semi-computable.
A function is called {\em semi-computable}
if it is either upper semi-computable or lower semi-computable.
If $f$ is both upper semi-computable and lower semi-computable,
then we call $f$ {\em computable} (or recursive if the domain
is integer or rational).

\end{definition}
Semi-computability gives no speed-of-convergence guaranties: even though
the limit value is monotonically approximated we know at no stage
in the process how close we are to the limit value.
The functions $h_x(\alpha), \lambda_x (\alpha), \beta_x(\alpha)$ have finite domain
for given $x$ and hence can be given as a table---so formally speaking
they are computable. But this evades the issue: there is no
algorithm that computes these functions for given $x$ and $\alpha$.   
Considering them as two-argument functions we
show the following (we actually quantify these):
\begin{itemize}
\item The functions $h_x(\alpha)$ and $\lambda_x(\alpha)$
are upper semi-computable but they are not computable up to any reasonable
precision.
\item Moreover, there is no algorithm that given $x^*$ and $\alpha$ finds 
$h_x(\alpha)$ or $\lambda_x(\alpha)$.
\item  The function $\beta_x(\alpha)$
is not upper- or lower semi-computable, not even to any reasonable precision,
but we can compute it given an oracle for the halting 
problem. 
\item There is no algorithm
that given $x$ and $K(x)$ finds a minimal sufficient statistic for $x$
up to any reasonable precision.
\end{itemize}

{\em Intuition: the functions $h_x$ and $\lambda_x$ (the ML-estimator
and the MDL-estimator, respectively) can be monotonically approximated
in the upper semi-computable sense. But the fitness function $\beta_x$
cannot be monotonically approximated in that sense, nor in the 
lower semi-computable sense, in both cases not even 
up to any relevant precision.
}

The precise forms of these quite strong noncomputability and
nonapproximability results are given in Appendix~\ref{app.comp}.

\section{Realizing the Structure Function}
\label{sect.real}

It is straightforward that we can {\em monotonically approximate} $h_x$
and its witnesses (similarly $\lambda_x$) in the sense that there
exists a non-halting algorithm $A$ that given any
$x,\alpha$ outputs a finite sequence $p_1,p_2,p_3,\dots,p_l$ of
pairwise different computer programs each of length at most
$\alpha+C\log|x|$ ($C$ is a constant) such that each program $p_i$ prints a 
model $S_i$ such that $|S_{1}|>|S_2| > \cdots > |S_l|$.
This way of computing $h_x$ or $\lambda_x$ is called upper semi-computable,
formally defined in Definition~\ref{def.semi}.  
By the results of Section~\ref{sect.coarse}
the last model $S_l$ is ``near'' the best possible model according
to the randomness deficiency criterion: There is no program
$p$ of length at most $\alpha$ that prints a model $S$ such that
the randomness deficiency of $x$ for $S$ is $C\log|x|$ less than
that of $x$ for $S_l$. Note that we are not
able to identify $p_l$ given $x,\alpha$, since the algorithm $A$ is
non-halting and thus we do not know which program will be
output last. 
This way we obtain a model of (approximately) best fit at 
each complexity level $\alpha$,
but non-uniformly.

The question arises whether there is a uniform construction
to obtain the models that realize the structure functions at given complexities.
Here we present such a construction. (In view of the non-computability
of structure functions, Section~\ref{sect.comput}, 
the construction is of course
not computable.)

We give a general uniform construction
of the finite sets witnessing $\lambda_x$, $h_x$, and $\beta_x$,
at each argument (that is, level of model complexity),
 in terms of
indexes of $x$ in the enumeration
of strings of given complexity, up to the ``coarse''
equivalence precision of Section~\ref{sect.coarse}. 
This extends a technique introduced
in \cite{GTV01}.

\begin{definition}\label{def.wx}
\rm
Let $N^l$ denote the number of strings of complexity at most $l$,
and  let $|N^l|$ denote the length of the binary notation of $N^l$.
For $i\le |N^l|$ let $N^l_i$ stand for $i$ most significant bits
of binary notation of $N^l$.
Let $D$ denote the set of all pairs $\{\pair{x,l}\mid K(x)\le l\}$.
Fix an enumeration of $D$ and denote by $I_x^l$
the minimum index of a pair $\pair{x,i}$ with $i\le l$
in that enumeration, that is,
the number of pairs enumerated before $\pair{x,i}$
(if $K(x)>l$ then
$I_x^l=\infty$).
Let $m_x^l$ denote the maximal common prefix of
binary notations of $I_x^l$ and $N^l$, that is, $I_x^l = m_x^l 0 ** \dots *$
and $N^l = m_x^l 1 ** \dots *$
(we assume here that binary notation
of $I_x^l$ is written in exactly $|N^l|$ bits with leading zeros if necessary).
\end{definition}
(In \cite{GTV01} the notation $m_x$ is used for $m_x^l$
with $l= K(x)$.)

	\begin{theorem}\label{th91}
For every  $i\le l$, the number	$N_i^l$
is algorithmically equivalent to $N^i$,
that is, $K(N^i \mid N_i^l),K(N_i^l \mid N^i)=O(\log l)$.
	\end{theorem}

Before proceeding to the main theorem of this section we
introduce some more notation.

\begin{definition}\label{def.mx}
\rm
For $i\le l$ let
$S_i^l$ denote the set
of all strings $y$ such that the binary notation of
$I^l_y$ has
the form $N^l_i0**\dots*$
(we assume here that binary notations
of indexes are written using exactly $|N^l|$ bits.)
\end{definition}

Let $c$ denote a constant such that
$K(x)\le \Lambda(S)+c$ for
every $x\in S$.
The following theorem shows that sets $S^l_i$ form a universal family of
statistics for $x$.

	\begin{theorem}\label{th90}
(i)
If the $(i+1)$st most significant bit of $N^l$ is $1$ then
$|S^l_i|=2^{|N^l|-i-1}$
and $S^l_i$ is algorithmically equivalent to $N^l_i$,
that is $K(N^l_i  \mid S^l_i),K(S^l_i  \mid N^l_i)=O(\log l)$.

(ii)
For every $S$ and every $x\in S$,
let $l=\Lambda(S)+c$ and  $i=|m^l_x|$.
Then  $x\in S^l_i$,
$K(S_i^l  \mid S)=O(\log l)$, $K(S_i^l)=i+O(\log l)\le K(S)+O(\log l)$,
and $\Lambda(S_i^l)\le \Lambda(S)+O(\log l)$
(that is, $S_i^l$ is not worse than $S$, as a model explaining $x$).

(iii)
If $\alpha$ is critical then
every $S$ witnessing
$\lambda_x(\alpha)$
is algorithmically equivalent to $N^{\alpha}$.
That is, if $K(S)\approx\alpha$ and $\Lambda(S)\approx\lambda_x(\alpha)$
but $K(N^\alpha|S)\gg 0$ or $K(S|N^\alpha)\gg 0$
then there is $A\ni x$ with
$K(A)\ll\alpha$ and $\Lambda(A)\approx\lambda_x(\alpha)$.
More specifically, for all $S\ni x$
either $K(S|N^\alpha)\le K(S)-\alpha$ and $K(N^\alpha|S)=0$,
or there is
$A\ni x$ such that
$\Lambda(A)\le\Lambda(S)$ and
$K(A)\le\min\{\alpha-K(N^\alpha|S),
K(S)-K(S|N^\alpha)\}$, where all inequalities hold
up to $O(\log\Lambda(S))$ additive term.

	\end{theorem}

Note that Item (iii) of the theorem does not hold for non-critical
points. For instance, for a random string
$x$ of length $n$ there are
independent $S_1,S_2$ witnessing $\lambda_x(\frac{n}{2})=n$:
let $S_1$ be the set of all $x'$ of length
$n$ having the same prefix of length $\frac{n}{2}$ as $x$ and
$S_2$ be the set of all $x'$ of length
$n$ having the same suffix of length $\frac{n}{2}$ as $x$.

\begin{corollary}
Let $x$ be a string of length $n$ and complexity $k$.
For every $\alpha$ ($K(n)+O(1)\le\alpha\le k$) there is $l\le n+K(n)+O(1)$ such
that the set $S^l_{\alpha}$
both contains $x$ and witnesses $h_x (\alpha)$,
$\lambda_x (\alpha)$,
and $\beta_x (\alpha)$,
up to an $O(\log n)$ additive term in the argument and value.
\end{corollary}

\appendix
\section{Oral History}\label{app.oral}
Since there is no written version of Kolmogorov's
initial proposal  \cite{Ko74b,Ko74a}, which we argued is a new approach to a
``non-probabilistic statistics,'' 
apart from a few lines \cite{Ko74b}
which we reproduced in Section \ref{sect.trans},
we have to rely on the testimony of witnesses
\cite{Ga02,Co02,Le02}.
Says Tom Cover \cite{Co02}: ``I remember
taking many long hours trying to understand the motivation of Kolmogorov's
approach.''
According to Peter G\'acs, \cite{Ga02}:
``Kolmogorov drew a picture of $h_x(\alpha)$ as
a function of $\alpha$ monotonically approaching the
diagonal [sufficiency line].
Kolmogorov stated that it was known
(proved by L.A. Levin) that
in some
cases
it remained far from
the diagonal line till the very end.''
 Leonid A. Levin \cite{Le02}:
``Kolmogorov told me  [about] $h_x(i)$ (or its inverse, I am not sure)
and asked how this $h(i)$ could behave. I proved that $i+h(i)+O(\log i)$
is monotone but otherwise arbitrary within $O(\sqrt{i})$ accuracy;
it stabilizes on $K(x)$ when $i$ exceeds $I(x : \text{Halting})$.
(Actually, this
expression for accuracy was Kolmogorov's re-wording, I gave it in less
elegant but equivalent terms---$O(p\log i)$ where p is the number of "jumps".)
I do not remember Kolmogorov defining  $\beta_x(i)$
or suggesting anything like your result.
I never published anything on the topic because I do not believe strings
$x$ with significant $I(x : \text{Halting})$ could exist in the world.''
($I(x:y) = K(y)-K(y|x)$ is the information 
in $x$ about $y$. By \eqref{eq.soi} we have $I(x:y)=I(y:x)$, with equality
holding up to a constant additive term indepennedent of $x$ and $y$,
and hence we call this quantity the {\em algorithmic mutual information}. 
Above,  "Halting" stands for the infinite binary ``halting sequence''
defined as follows:
The $i$th bit of Halting is 1 iff the $i$th program for the reference universal
prefix machine $U$ halts, and 0 otherwise.) 

\begin{remark}
\rm
Levin's statement \cite{Le02} quoted above appears to suggest that
strings $x$ such that $h_x(i)+i$ stabilizes on $K(x)$ only for
large $i$ may exist mathematically but are unlikely to occur
in nature, because such $x$'s must have a lot of 
information about the Halting problem. 
and hence the analysis of their properties is irrelevant.
But the statement in question is imprecise.
There are two ways to understand the statement:
(i) $h_x(i)+i$ stabilizes on $K(x)$ when $i$ exceeds $I(x : \text{Halting})$
          or earlier; or
(ii) $h_x(i)+i$ stabilizes on $K(x)$ when $i$ exceeds $I(x : \text{Halting})$
          and not earlier.
It is not clear what 
``the information in $x$ about the halting problem'' is, since
the ``Halting problem'' is not a finite object and thus the notion of
information about Halting needs a special definition.
The usual $I(x:\text{Halting})=K(\text{Halting})-K(\text{Halting} \mid x)$ 
doesn't make sense since
both 
$K(\text{Halting})$ and $K(\text{Halting} \mid x)$ are infinite.
The expression $I(x:\text{Halting})=K(x)-K(x \mid \text{Halting})$ looks better
provided $K(x \mid \text{Halting})$ is understood as $K(x)$ relativized by the
Halting problem.
In the latter interpretation of $I(x:\text{Halting})$,
case (i) is correct and case (ii) is false.
The correctness of (i) is implicit in Theorem V.4.
A counter example to (ii):
Let $p$ be the halting program of length
at most $n$ with the greatest running time.
It is easy to show that $K(p)$ is about $n$, and
therefore $p$ is a random string of length about $n$.
As a consequence, the complexity of the
 minimal sufficient statistic $\alpha_0$
of $p$ is close to 0.
On the other hand $I(p:\text{Halting})$ is about $n$.
Indeed, given the oracle for the Halting problem and $n$ we can find $x$;
hence $I(p:\text{Halting}) = 
K(p) - K(p \mid \text{Halting}) \ge n - K(n) \ge n - 2\log n$.
\end{remark}

\section{Validity for Extended Models}\label{app.extension}
\begin{figure}
\begin{center}
\epsfxsize=8cm
\epsfxsize=8cm \epsfbox{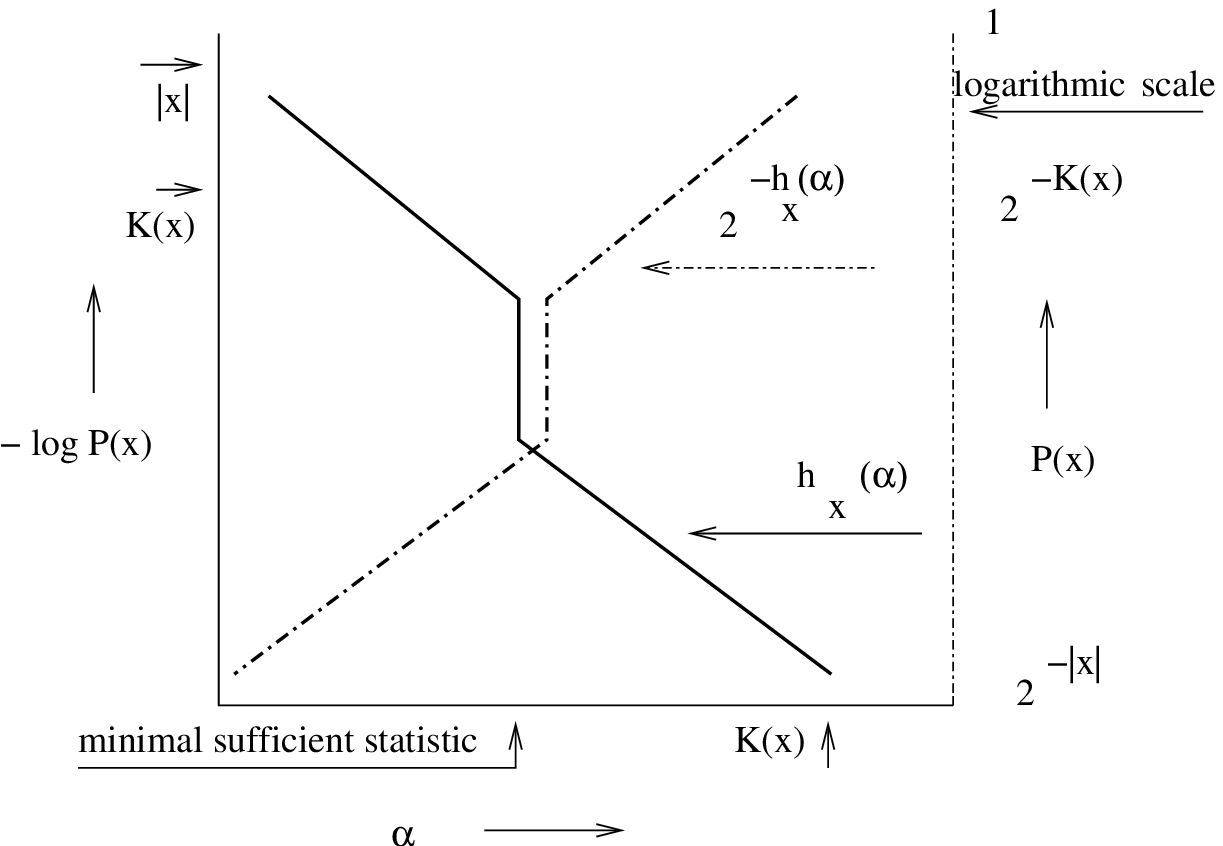}
\end{center}
\caption{Structure function $h_x( \alpha)= \min_P \{- \log P(x): P(x)>0, \; K(P) \leq
 \alpha\}$ with $P$ a computable
probability density function, with values according to the left
vertical coordinate, and the maximum likelihood estimator $2^{-h_x(\alpha)}=
\max \{P(x): P(x)>0 , \; K(P) \leq \alpha\}$,
with values according to the right-hand side vertical coordinate.}
\label{figure.MLestimator}
\end{figure}

Following Kolmogorov we analyzed a canonical setting
where the models are finite sets. As Kolmogorov himself pointed
out, this is no real restriction:
the finite sets model class is equivalent, up to a logarithmic additive
term, to the model class of probability density functions, as
studied in \cite{Sh83,GTV01}.
The analysis is valid, up to logarithmic
additive terms, also for
the model class of total recursive functions, as studied in \cite{Vi01}.
The model class of {\em computable probability
density functions} consists of the set
of functions $P: \{0,1\}^* \rightarrow [0,1]$ with
$\sum P(x) = 1$.
``Computable'' means here that there is a Turing machine $T_P$ that,
given $x$ and a positive rational  $\eps$,
computes $P(x)$ with precision $\eps$.
The (prefix-) complexity $K(P)$ of a
computable (possibly partial) function $P$ is defined by
$
K(P) = \min_i \{K(i): \mbox{\rm Turing machine } T_i
\; \; \mbox{\rm computes }
P \}.
$
A string $x$ is typical for a distribution $P$ if the randomness
deficiency
$
\delta (x \mid P) =  - \log P(x) - K(x \mid P)
$
is small. The conditional complexity
$K(x \mid P)$ is defined as follows. Say that a function
$A$ approximates $P$ if $|A(y,\eps)-P(y)|<\eps$
for every $y$ and every positive rational
$\eps$. Then $K(x \mid P)$ is
the minimum length
of a program that given every function $A$ approximating $P$
as an oracle prints $x$.
Similarly, $P$ is $c$-optimal for $x$ if
$
K(P)  - \log P(x) \leq K(x)+c
$.
Thus, instead of the data-to-model code length $\log|S|$ for
finite set models, we consider the data-to-model code length
$-\log P(x)$ (the Shannon-Fano code). The value $-\log P(x)$
measures also how likely $x$ is under the hypothesis $P$ and the mapping
$x\mapsto P_{\min}$ where $P_{\min}$ minimizes $-\log P(x)$ over $P$ with 
$K(P)\le\alpha$ is a \emph{constrained maximum likelihood estimator},
see Figure~\ref{figure.MLestimator}. Our
results thus imply that such a constrained maximum likelihood estimator always
returns a hypothesis with minimum randomness deficiency.

The essence of this approach is that we mean maximization 
over a class of likelihoods induced by
computable probability density functions 
that are below a certain complexity level $\alpha$.
In classical statistics, unconstrained maximal likelihood is known
to perform badly for model selection, because
it tends to want the most complex models possible. This is
closely reflected in our approach: unconstrained maximization
will result in the computable probability distribution of complexity
about $K(x)$ that
concentrates all probability on $x$. But the structure function 
$h_x(\alpha)$ tells us all stochastic properties of data $x$ in
the sense as explained in detail in the start of Section~\ref{sect.coarse}.
for finite set models.

The model class of {\em total recursive
functions} consists of the set
of computable functions $p: \{0,1\}^* \rightarrow \{0,1\}^*$.
The (prefix-) complexity $K(p)$ of a
total recursive function $p$ is defined by
$
K(p) = \min_i \{K(i): \mbox{\rm Turing machine } T_i
\; \; \mbox{\rm computes }
p \}.
$
In place of $\log|S|$ for finite set models
we consider the data-to-model code length
$\len xp=\min\{|d|:p(d)=x\}$.
A string $x$ is typical for a total recursive function $p$ if
the randomness
deficiency
$
\delta (x \mid p) = \len xp- K(x \mid p)
$
is small.
The conditional complexity
$K(x \mid p)$ is defined as
the minimum length
of a program that given $p$
as an oracle prints $x$.
Similarly, $p$ is $c$-optimal for $x$ if
$
K(p)  + \len xp \leq K(x)+c . 
$

It is easy to show that for every data string $x$
and a contemplated finite set model for it, there
is an almost equivalent computable probability density function
model and an almost equivalent total recursive function model.
 
\begin{proposition}\label{prop.1}
For every $x$ and every finite set $S \ni x$ there is:

(a) A computable probability
density function $P$ with $-\log P(x) =\log|S|$,
$\delta(x \mid P)=\delta(x \mid S)+O(1)$
and $K(P) = K(S)+ O(1)$; and

(b) A total recursive function $p$
such that $\len xp\le\log|S|$,
$\delta(x \mid p)\le\delta(x \mid S)+O(1)$
and
$K(p) = K(S)+ O(1)$. 
\end{proposition}

\begin{proof}
(a) Define $P(y)= 1/|S|$ for $y \in S$ 
and 0 otherwise.

(b) If $S = \{x_0, \ldots , x_{m-1}\}$, then define $p(d) =
x_{d \bmod m}$.
\end{proof} 

The converse of Proposition~\ref{prop.1} is slightly harder:
for every data string $x$ and a contemplated
computable probability density function model for it,
as well as for a contemplated total recursive function model for $x$,
there is a finite set model for $x$ that has no worse complexity,
randomness deficiency, and worst-case data-to-model code for $x$,
up to additive logarithmic precision.

    \begin{proposition}\label{prop.2}
    There are constants $c,C$, such that
    for every string $x$, the following holds:

    (a) For every computable probability
    density function $P$
    there is a finite set $S \ni x$
    such that $\log |S| < - \log P(x)+1$, $\delta(x \mid S)\le \delta(x \mid P)+
    2\log K(P)+K(\lfloor-\log P(x)\rfloor)+2\log K(\lfloor-\log P(x)\rfloor)+C$
    and
    $K(S) \leq  K(P) + K(\lfloor-\log P(x)\rfloor)+C$; and

    (b) For every total recursive function
    $p$
    there is a finite set $S \ni x$ with $\log |S| \leq \len xp$,
    $\delta(x \mid S)\le \delta(x \mid p)+
    2\log K(p)+K(\len xp)+2\log K(\len xp)+c$
    and
    $K(S) \leq  K(p) + K(\len xp)+c$.
    \end{proposition}

\begin{proof}
(a)
Let $m=\lfloor-\log P(x)\rfloor$, that is,
$2^{-m-1}<P(x)\le 2^{-m}$.
Define $S = \{y:  P(y)
> 2^{-m-1}\}$. Then,
$|S|<2^{m+1} \leq 2/P(x)$,
which implies the claimed value for $\log |S|$.
To list $S$ it suffices to compute all consecutive values of $P(y)$ to
sufficient precision
until the combined probabilities exceed $1-2^{-m-1}$.
That is, $K(S) \leq
K(P)+ K(m)+O(1)$.
Finally,
$\delta(x \mid S)=\log|S|-K(x|S^*)<-\log P(x)-K(x \mid S^*)+1=
\delta(x \mid P)+K(x \mid P)-K(x \mid S^*)+1\le \delta(x \mid P)+K(S^* \mid P)+O(1)$.
The term $K(S^* \mid P)$ can be upper bounded
as $K(K(S))+K(m)+O(1)\le 2\log K(S)+K(m)+O(1)
\le 2\log (K(P)+K(m))+K(m)+O(1)
\le 2\log K(P)+2\log K(m)+K(m)+O(1)$, which implies the claimed bound for
 $\delta(x \mid S)$.

(b) Define $S = \{y: p(d) = y, \; \; |d| = \len xp  \}$.
Then,
$\log |S| \leq \len xp$. To list $S$ it suffices to compute $p(d)$ for
every argument of length equal $\len xp$. Hence,
$K(S) \leq K(p)+K(\len xp)+O(1)$.
The upper bound for $\delta(x \mid S)$ is derived just in the same way as
in the proof of item (a).
\end{proof}

\begin{remark}
\rm
How large are the nonconstant additive complexity terms in
Proposition~\ref{prop.2} for strings $x$ of length $n$? In item (a),
we are commonly only interested 
in $P$ such that $K(P)\le n+O(\log n)$ and
$-\log P(x)\le n+O(1)$.
Indeed, for every $P$ there is $P'$ such that
$K(P')\le \min\{K(P),n\}+O(\log n)$,
$\delta(x \mid P')\le \delta(x \mid P)+O(\log n)$,
$-\log P'(x)\le\min\{-\log P(x),n\}+1$.
Such $P'$ is defined as follows: If
$K(P)>n$ then $P'(x)=1$ and $P'(y)=0$ for every $y\ne x$;
otherwise $P'=(P+U_n)/2$ where $U_n$ stands for
the uniform distribution
on $\{0,1\}^n$.
Then the additive terms in item (a) are $O(\log n)$.
In item (b) we are commonly only interested in $p$ such that
$K(p)\le n+O(\log n)$ and $\len xp\le n+O(1)$.
Indeed, for every $p$  there is $p'$
such that
$K(p')\le \min\{K(p),n\}+O(\log n)$,
$\delta(x \mid p')\le \delta(x \mid p)+O(\log n)$,
$\len x{p'}\le\min\{\len xp,n\}+1$.
Such $p'$ is defined as follows: If
$K(p)>n$ then $p'$ maps all strings to $x$;
otherwise
$p'(0u)=p(u)$ and
$p'(1u)=u$.
Then the additive terms in item (b) are $O(\log n)$.
Thus, in this sense all results in this paper that hold for
finite set models extend, up to a logarithmic
additive term, to computable probability density function models
and to total recursive function models. Since the results in
this paper hold only up to additive logarithmic term anyway, this means
that all of them equivalently hold for the model
class of computable probability density 
functions, as well as for the model class of total recursive functions.
\end{remark}

\section{Proofs}\label{app.proof}
      \begin{proof} {\bf Lemma~\ref{lem.equiv}}
The inequality
$\lambda_x(\alpha)\le h_x(\alpha)+\alpha$ is immediate.
So it suffices to prove that
$h_x(\alpha)+\alpha\le \lambda_x(\alpha)+K(\alpha)+O(1)$.
The proof of this inequality is based on the following:

	   \begin{claim}
Ignoring additive $K(i)$ terms the function
$h_x(i)+i$ does not increase:
		\begin{equation}\label{ineq1}
h_x(i_2)+i_2\le h_x(i_1)+i_1+K(i_2 \mid i_1)+ O(1)
		\end{equation}
for $i_1<i_2\le K(x)$.
	   \end{claim}
	\begin{proof}
Let $S$ be a finite set containing $x$ with $K(S) \leq i_1$ and 
$\log |S| = h_x (i_1)$. For every
$m\le\log|S|$, we can partition $S$ into
$2^m$ equal-size parts and select the part $S'$ containing $x$.
Then, $\log |S'| = \log|S|-m$
at the cost of increasing
the complexity of $S'$ to
	   $$
K(S') \leq K(S)+m+K(m  \mid K(S))+O(1)
	   $$
(we specify the part $S'$ containing
$x$ by its index among all the parts).
Choose
	   $$
m=i_2-K(S)- K(i_2  \mid K(S)) - c
	   $$
for a constant $c$
to be determined later.
Note that
\begin{align*}
K(m  \mid K(S)) & \leq
K(i_2,K(i_2  \mid K(S))  \mid K(S))+K(c)+c'
\\ &=
K(i_2  \mid K(S))+K(c)+c''
\end{align*}
for appropriate constants $c',c''$.
The complexity of the resulting set $S'$ is thus at most
\begin{align*}
K(S) +  i_2-K(S)& - K(i_2  \mid K(S)) - c \\
& +K(i_2  \mid K(S))+K(c)+c'' \leq i_2,
\end{align*}
provided $c$ is chosen
large enough. Hence,
$h_x (i_2 ) \leq \log |S'| = h_x (i_1) - m=h_x (i_1) -
i_2+K(S)+K(i_2  \mid K(S)) +c$, and it suffices to prove that
$K(S)+K(i_2  \mid K(S))\le i_1+K(i_2  \mid i_1)+O(1)$.
This follows from the bound
$K(i_2  \mid K(S))\leq K(i_2  \mid i_1)+K(i_1  \mid K(S))+O(1)
\leq K(i_2  \mid i_1)+K(i_1-K(S))+O(1)
\le K(i_2 \mid i_1)+i_1-K(S)+O(1)$.
	\end{proof}

Let $S$ witness $\lambda_x(\alpha)$.
Substituting $K(S)=i_1$, $\alpha = i_2$  in  \eqref{ineq1}
we obtain:
		$
h_x(\alpha)+\alpha\le h_x(K(S))+K(S)+K(\alpha  \mid K(S))+ O(1)
\le \Lambda(S) +K( \alpha ) + O(1)
 = \lambda_x(\alpha)+K(\alpha)+ O(1).
		$
\end{proof}

           \begin{proof} {\bf Theorem~\ref{th14}}
(i)
We first observe that for every $x$ of length $n$ we have
$\lambda_x(K(n)+O(1))\leq n + K(n)+O(1)$,
as witnessed by $S=\{0,1\}^n$.
At the other extreme,
$\lambda_x(k+O(1))= k+O(1)$, as witnessed by
$S=\{x\}$.

Define $\lambda(i)$ by the equation
$\lambda(i)-k=\max\{0,\lambda_x(i+K(n)+O(1))-k-O(1)\}$.
Then $\close {\lambda_x}\lambda$ with $\eps=\delta = K(n)+O(1)$,
and $\lambda$ satisfies the requirements
of Item (i) of the theorem.

(ii) Fix $\lambda(i)$ satisfying the conditions in the theorem.
It suffices to show that
there is a string $x$ of length $n$ such that, for every 
$i \in [0,k]$, we have
$\lambda_x(i)\ge \lambda(i)$ and
$\lambda_x(i+\delta (i))\le \lambda(i)+\delta (i)$ for
$\delta (i) = K(i,n,\lambda)+O(1)$.
Then, with $\delta=\delta(k)$, 
we have 
$K(x)\le \lambda_x(k+\delta)+O(1)\le
\lambda(k)+\delta+O(1)=k+\delta+O(1)$.
And the inequality
$\lambda_x(k)\ge \lambda(k)= k$
implies that
$K(x)>k-O(1)$.

\begin{claim}\label{claim.simult}
\rm
For every length $n$,
there is a string $x$ of length $n$ such that
$\lambda_x( i ) \ge \lambda( i)$
for every  $i$ in the domain of $\lambda$.
\end{claim}
\begin{proof}
Fix a length $n$.
If
$\lambda_x( i ) < \lambda( i)$
then $x$ belongs to a set $A$ with
$\Lambda(A)<\lambda(i)
\le\lambda(0)\le n$.
The total number of elements  in different such $A$'s
is less than $\sum_A2^{n-K(A)}= 2^n\sum_A2^{-K(A)}\le 2^n$, where the
second inequality follows by (\ref{eq.udconv}).
\end{proof}

We prove Item (ii) by demonstrating
that the lexicographically first $x$, as defined in
Claim~\ref{claim.simult},
also satisfies
$\lambda_x(i +\delta(i))\le \lambda(i)+\delta(i)$,
for $\delta(i) = K(i,n,\lambda)+O(1)$
for all $i \in [0,k]$.
It suffices to construct a set $S \ni x$
of cardinality $2^{\lambda(i)-i}$
and of complexity at most $i +\delta (i)$,  
for every $i \in [0,  k]$.

For every fixed $i \in [0,k]$ we can run the following:

{\bf Algorithm:} 
Let $A$ be a set variable initially containing all strings of length $n$, 
and let $S$ be a set variable initially containing
the $2^{\lambda(i)-i}$ first strings of $A$ in lexicographical order.
Run all programs of length at most $n$
dovetail style.  Every time 
a program $p$ of some length
$j$ halts, $\lambda(j)$ is defined,  and $p$
prints a set $B$ of cardinality
at most $2^{\lambda(j)-j}$, we remove all the elements of $B$
from $A$ (but not from $S$); we
call a step at which this happens a {\em $j$-step}.
Every time $S\cap A$ becomes empty at a $j$-step, 
we replace the contents of $S$ by the set of  the
$2^{\lambda(i)-i}$ first strings in lexicographical order of
(the current contents of) $A$. 
Possibly, the last replacement
of $S$ is incomplete because there are less than $2^{\lambda(i)-i}$
elements left in $A$.
It is easy to see that $x \in S \setminus A$ 
just after the final replacement, 
and stays there forever after, even though 
some programs in the dovetailing process may still be running and 
elements from $A$ may still be eliminated.

\begin{claim}\label{claim.change}
\rm
The contents of the set $S$ is replaced at most $2^{i+1}$ times.
\end{claim}

\begin{proof}
There are two types of replacements that will be treated separately.

{\bf Case 1:} Replacement of the current contents of $S$ where 
at some $j$-step with $j\le i$ {\em at least one} element was removed
from the current contents $S \cap A$.
Trivially, the number of this type of replacements
is bounded by the number of $j$-steps with $j< i$, and hence by the number
of programs of length less than $i$, that is, by
$2^{i}$.

{\bf Case 2:}  Replacement of the current contents of $S$ 
where {\em every one} of the $2^{\lambda(i)-i}$
elements of the current contents of $S$
is removed from $A$ by $j$-steps with $j\ge i$.
Let us estimate the number of this type of replacements:
Every element $x$ removed at a $j$-step with $j\ge i$ belongs
to a set $B$ with $\Lambda(B)\le
\lambda(j)\le\lambda(i)$.
The overall cumulative number of elements removed from $A$ on
$j$-steps with $j\ge i$ is bounded by
$\sum_B2^{\lambda(i)-K(B)}\le2^{\lambda(i)}$, where the
inequality follows by (\ref{eq.udconv}).
Hence replacements of the second type can happen at most
$2^{\lambda(i)-(\lambda(i)-i)}=2^{i}$ times.
\end{proof}

By Claim~\ref{claim.change},
$S$ stabilizes after a certain number of $j$-steps. 
That number may be large. However, the number of replacements
of $S$ is small.
The final set $S \ni x$
has cardinality $2^{\lambda(i)-i}$, and can be
specified by the number of replacements resulting in its current
contents (as in Claim~\ref{claim.change}), 
and by $i,n,\lambda$. This shows that
$K(S) \le i+K(i,n,\lambda)+O(1)$.
	  \end{proof}

           \begin{proof} {\bf Theorem~\ref{th13}}
   The statement of the theorem easily follows from the following two
   inequalities that are valid for every $x$
   (where $n=|x|$ and $k=K(x)$):
\begin{align}\label{eq.i}
& \beta_x(i)+k\le \lambda_x(i)+O(1), \mbox{\rm \;  for every
$i\le k$;  and}
\\ \label{eq.ii}
& \lambda_x(i+O(\log n))\le \beta_x(i)+k+O(\log n), \\
&\mbox{\rm  \; for every $i$ satisfying 
$K(n)+O(1) \le i \le k$.}
\nonumber
\end{align}
It is convenient to rewrite the formula defining
$\delta(x \mid A)$
using the symmetry of information \eqref{eq.soi} as follows:
\begin{align}\label{eq84}
 \delta (x \mid A)& =\log|A|+K(A)-K(A \mid x^*)-k+O(1)
\\& =
\Lambda(A)-K(A \mid x^*)-k+O(1).
\nonumber
\end{align}

Ad \eqref{eq.i}: This is easy, because
for every set $S \ni x$ witnessing $\lambda_x(i)$
we have
$\delta (x \mid S)\le\Lambda(S)-k+O(1)=
\lambda_x(i)-k+O(1)$
and $\beta_x (i) \leq \delta (x \mid S)$.

    Ad \eqref{eq.ii}: This is more difficult.
    By \eqref{eq84}, and the obvious
    $K(A \mid x^*)\leq K(A \mid x)+O(1)$,
it suffices to prove
that for every $A\ni x$ there is an $S\ni x$ with
        \begin{align*}
&K(S)\le K(A) +O(\log m),\\
&\log|S|\le\log|A| -K(A \mid x)+O(\log m),
        \end{align*}
where $m=\Lambda(A)$.
Indeed for every $A$ witnessing
$\beta_x(i)$ the set $S$ will witness
$\lambda_x(i+O(\log n))\le \beta_x(i)+k+O(\log n)$
(note that $m=
\log|A|+K(A)=K(x\mid A^*)+\beta_x(i)+K(A)
\le 3n+O(\log n)$ provided $i\ge K(n)+O(1)$).
The above assertion is only a little bit easier  to  prove  than
the one in Lemma~\ref{l2} below that also suffices.
Since we need this lemma in any case in the proof of  Theorem~\ref{thsxa}
we state and prove it right now.

        \begin{lemma}\label{l2}
For every $A\ni x$ there is $S\ni x$ with
$K(S)\le K(A)-K(A \mid x)+O(\log m)$ and
$\lceil\log|S|\rceil=\lceil\log|A|\rceil$ {\em (}where $m=\Lambda (A)$
{\em )}.
        \end{lemma}

\begin{proof}
Fix some $A_0 \ni x$ and let $m = \Lambda (A_0)$.
Our task is the following:  Given $K(A_0),\lceil\log|A_0|\rceil, K(A_0 \mid x)$,
to enumerate a family of at most $2^{K(A_0) - K(A_0 \mid x) +O(\log m)}$
different sets $S$ with $\log |S| = \lceil\log|A_0|\rceil$
that cover
all $y$'s covered by sets $A$, with
$K(A)=K(A_0)$,
$K(A \mid y)=K(A_0 \mid x)$ and
$\lceil\log|A|\rceil=\lceil\log|A_0|\rceil$.
Since
the complexity of each enumerated $S$ does not exceed
$K(A_0)-K(A_0 \mid x)+O(\log m)+K(K(A_0),\lceil\log|A_0|\rceil,K(A_0 \mid x)) + O(1)=
K(A_0)-K(A_0 \mid x)+O(\log m)$ the lemma
will be proved.
The proof is by running the following:

{\bf Algorithm:} 
Given $K(A_0),\lceil\log|A_0|\rceil,K(A_0  \mid x)$
we run all programs
dovetail style.
We maintain auxiliary set-variables $C,U,D$, all of them
initially $\emptyset$.
Every time a new program $p$ of length $K(A_0)$
in the dovetailing process halts, with as
output a set $A$ with $\lceil\log|A|\rceil = \lceil\log|A_0|\rceil$,
we execute the following steps:
\begin{description}
\item[{\bf Step 1:}]
Update $U := U \cup A$.
\item[{\bf Step 2:}]
Update $D := \{y \in U\setminus C$: $y$ is
covered by at least $t= 2^{K(A_0 \mid x)-\delta}$
different  generated $A$'s$\}$,
where $\delta=O(\log m)$ will be defined later. 
\item[{\bf Step 3:}]
This
step is executed only if there is $y\in D$ that is covered
by at least $2t$ different  generated $A$'s.
Enumerate as much new disjoint sets $S$ as are needed to
cover $D$: we just chop $D$ into parts
of size $2^{\lceil\log|A_0|\rceil}$ (the last part may be incomplete)
and name those parts the new  sets $S$.
Every time a new set $S$ is enumerated, update
$C := C \cup S$.
\end{description}

\begin{claim}
\rm
The string $x$ is an element of some enumerated $S$, and
the number of enumerated $S$'s is
at most $2^{K(A_0)-K(A_0  \mid x)+ O(\log m)}$.
\end{claim}

\begin{proof}
By way of contradiction,
assume that $x$ is not an element of the enumerated $S$'s. 
Then there are less than $2^{K(A_0 \mid x)-\delta+1}$
different generated sets $A$ such that
$x\in A$. Every such $A$ therefore satisfies
$K(A\mid x)\le K(A_0 \mid x)-\delta+O(\log m)<K(A_0 \mid x)$
if $\delta$ is chosen appropriately.
Since $A_0$ was certainly generated this is a contradiction.


It remains to
show that we enumerated at most $2^{K(A_0)-K(A_0 \mid x)+O(\log m)}$
different $S$'s.
Step 3 is executed only once per $t$ executions of Step 1,
and Step 1 is executed at most
$2^{K(A_0)}$ times.
Therefore Step 3 is executed at most
$2^{K(A_0)}/t=2^{K(A_0)-K(A_0 \mid x)+\delta}$ times.
The number of $S$'s formed from incomplete parts of
$D$'s in Step 3 is thus at most $2^{K(A_0)-K(A_0)+\delta}$.
Let us bound
the number of $S$'s formed from complete parts of
$D$'s.
The total number of elements  in different $A$'s generated
is at most $2^{K(A_0)+\lceil\log|A_0|\rceil}$
counting multiplicity. Therefore the number of
elements in their union,
having multiplicity $2^{K(A_0 \mid x)-\delta}$
or more, is at most $2^{K(A_0)+\lceil\log|A_0|\rceil-K(A_0  \mid x)+\delta}$.
Every $S$ formed from a
complete part of a set $D$ in Step 3 accounts for
$2^{\lceil\log|A_0|\rceil}$ of them. Hence the number of $S$'s formed from
complete
parts of $D$'s is at most  $2^{K(A_0)-K(A_0 \mid x)+\delta}$.
\end{proof}
\end{proof}
\end{proof}

        \begin{proof} {\bf Theorem~\ref{thsxa}}
By Lemma~\ref{l2} there is $S\ni x$ with
$K(S)\le K(A)-K(A \mid x)+O(\log \Lambda(A))$ and
$\lceil\log|S|\rceil=\lceil\log|A|\rceil$.

Let us upper bound first
$K(S)$. We have
              \begin{align*}
K(S)
&\le K(A)-K(A \mid x)
=\delta(x|A)+k-\log|A|\\
&=\beta_x(\alpha)+k-\log|A|+(\delta(x|A)-\beta_x(\alpha))\\
&\le\lambda_x(\alpha)-\log|A|+(\delta(x|A)-\beta_x(\alpha)).
              \end{align*}
(all inequalities are valid up to $O(\log \Lambda(A))$
additive term).
The obtained upper bound is obviosly equivalent
to the first upper bound of $K(S)$ in the theorem.
As $\log|S|=\log|A|$ it gives the upper bound
of $\Lambda(S)$ from the theorem.
Finally, as  $\lambda_x(\alpha)\le h_x(\alpha)+\alpha+O(1)$
we obtain
$K(S)\le \alpha+ (h_x(\alpha)-\log|A|)+(\delta(x|A)-\beta_x(\alpha))$
(up to $O(\log \Lambda(A))$ additive term).
        \end{proof}

	\begin{proof} {\bf Theorem~\ref{th91}}
We first show that $|m_x^l|\le K(x)+O(\log l)$ for every
$x$ with $K(x)\le l$.
Indeed, given $x$, $l$, $|m_x^l|$ and the
$|N^l|-|m_x^l|$ least significant bits of $N^l$ we can find $N^l$: find
$I_x^l$ by enumerating $D$ until a pair $\pair{x,i}$ with $i\le l$
appears and then
complete $m_x^l$ by using
the $|m_x^l|$ most significant bits of the binary representation of $I_x^l$.
Given $l$ and $N^l$ we can find, using a constant-length
program, the lexicographically first string not in $N^l$. By construction,
this string 
has complexity at
least $l+1$. Then, $l\le K(N^l)+O(\log l)\le K(x)+|N^l|-|m_x^l|+O(\log l)
\le K(x)+l-|m_x^l|+O(\log l)$ (use $|N^l| \leq l+O(1)$).
Thus, $|m_x^l|\le K(x)+O(\log l)$.

Let $x$ be the string of complexity at most $i$ with maximum $I_x^l$.
Given $m_x^l$ and $i,l,|N^l|$ we can find
all strings of complexity at most $i$
by  enumerating $D$ until $N$ pairs $\pair{y,j}$
with $j\le l$ appear, where
$N$ is the number whose binary representation has
prefix $m^l_x1$ and then $(|N^l|-|m^l_x|-1)$ zeros.
Since  $|m^l_x|\le i+O(\log l)$,
this proves $K(N^i  \mid N^l_i)=O(\log l)$.
Since $K(N^i)\ge i-O(\log i)\ge K(N^l_i)-O(\log i)$ we have
$K(N^l_i  \mid N^i)=O(\log l)$.
	\end{proof}

	\begin{proof} {\bf Theorem~\ref{th90}}
(i)
If the $(i+1)$st most significant bit of $N^l$ is ``1,'' then
all the numbers with binary representation of
the form $N^l_i0**\dots*$ are used as indexes of some
$y$ with $K(y)\le l$, that is,
$S_i^l$ has exactly $2^{|N^l|-i-1}$ elements.
We can find $S_i^l$  given $l$, $i$, $|N^l|$ and
$N^l_i$ by enumerating all its elements.
On the other hand, $N^l_i$ can be found given
$S_i^l$ and $i,l$ as the first $i$ bits
of $I^l_x$ for every  $x\in S^l_i$.

    (ii) Since $i=|m^l_x|$, the largest
    common prefix of binary representation of $I^l_x$ and $N^l$
    has the form $N^l_i0**\dots*$ and
the $(i+1)$st most significant bit of $N^l$ is 1.
In particular, $x\in S^l_i$.

Let $J=\max\{I^l_y\mid y\in S\}$.
As $x\in S$, we have $J\ge I^l_x$.
We can find $N^l_i$ given $i$, $l$ and $S$ by finding
$J$ and taking the $i$ first bits of $J$.
Given $N^l_i$ we can find $S_i^l$. Hence
$K(S^l_i  \mid S)=O(\log l)$.
Therefore  $K(S_i^l)\le K(S)+O(\log l)$.
By Item (i) and by previous theorem we have
$K(S_i^l)=i+O(\log l)$.
Again by Item (i)
we have $\Lambda(S_i^l)\le l+O(\log l)=\Lambda(S)+O(\log l)$.

(iii) Let
$i=|m^l_x|$.
We distinguish two cases.

{\bf Case 1:} $i\ge\alpha$.
Then $K(N^\alpha  \mid S)\le K(N^\alpha  \mid S^l_i)+O(\log l)
\le K(N^\alpha  \mid N^i)+O(\log l)=O(\log l)$.
And
$K(S  \mid N^\alpha)=K(S)-K(N^\alpha)+O(\log l)=K(S)-\alpha+O(\log l)$.

{\bf Case 2:} $i<\alpha$. Let $A=S^l_i$.
As $\Lambda(S_i^l)\le \Lambda(S)+O(\log l)$ we need
to prove that
$K(S_i^l)\le\alpha-K(N^\alpha|S)$ and
$K(S_i^l)\le K(S)-K(S|N^\alpha)$
up to $O(\log l)$ additive term.
We have
              \begin{align*}
K(S^l_i) & =  K(N^\alpha)-K(N^\alpha|S^l_i)+O(\log l)
\\& \le
\alpha  - K(N^\alpha|S)+O(\log l)
              \end{align*}
and
              \begin{align*}
K(S^l_i) & =K(S)-K(S|S^l_i) +O(\log l)
\\& \le  K(S)-K(S|N^\alpha) +O(\log l).
              \end{align*}

	\end{proof}

\section{Computability properties}
\label{app.comp}
\subsection{Structure Function}
It is easy to see that $h_x(\alpha)$ or $\lambda_x (\alpha)$, and the
finite set that witnesses its value, are
upper semi-computable: run all programs of length up to $\alpha$
dovetailed fashion,
check whether a halting program produced a finite set containing
$x$, and replace the previous candidate with the new set if it is
smaller.

The next question is: 
Is the function 
$\lambda_x(\alpha)$, as the function of two arguments, computable?
Of course not, because if this were the case, then we 
could find, given every large $k$, a string 
of complexity at least $k$. 
Indeed, we know that there is a string $x$ 
for which $\lambda_x(k)>k$.
Applying the algorithm to all strings in the lexicographical order
find the first such $x$. Obviously $K(x)\ge k-O(1)$.
But it is known that we cannot prove that $K(x) > k$
for sufficiently large $k$, \cite{LiVi97}.

Assume now that we are given also
$K(x)$. The above argument does not work any more but the statement
remains true: $\lambda_x(\alpha)$ is not computable even if the algorithm is
given $K(x)$. 

Assume first that the algorithm is required to output the correct answer
given any approximation to $K(x)$. We show
that
no algorithm can find $\lambda$
that is close to $\lambda_x(\alpha)$
for some $0\ll\alpha\ll K(x)$.

	\begin{theorem}\label{theo.th6}
For every constant $c$ there is a constant $d$
such the following holds.
There is no  algorithm that
for infinitely many $k$, given $k$ and $x$ of length $k+d\log k$
with $|K(x)-k|\le 2\log k$, always 
finds $\lambda$ such that there is
$2\log k\le\alpha\le k$
with
$|\lambda_x(\alpha)-\lambda|<c\log k$.
	\end{theorem}
	\begin{proof}
Fix $c$. The value of $d$ will be chosen later.
The proof is by contradiction. 
Let $A$ be some algorithm.
We want to fool it on some pair $\pair{x,k}$.

Fix large $k$. 
We will construct a set $S$ of cardinality
$2^{k-2\log k}$ such that every string $x$ in $S$
has length $k+d\log k$ and complexity at most 
$k+2\log k$, and 
the algorithm halts on $\pair{x,k}$
and outputs $\lambda\ge (c+1)\log k$.
This is a contradiction.
Indeed, there is $x\in S$ with $K(x)\ge k-2\log k$. Hence
the output  $\lambda$ of $A$ on $\pair{x,k}$ is correct, that is,
there is $\alpha$ with $2\log k\le\alpha\le k$
and $|\lambda_x(\alpha)-\lambda|<c\log k$.
Then $\lambda_x(\alpha)>\log k$.
On the other hand, $\lambda(2\log k)\le k$ as witnessed
by $S$. Thus we obtain
	$$
k<\lambda_x(\alpha)\le \lambda_x(2\log k)\le k,
	$$ 
a contradiction.

Run in a dovetailed fashion all programs of length $k$ or less.
Start with $x$ equal to the first string of length $k+d\log k$ and with
$S=B=\emptyset$.
Run $A$ on $\pair{x,k}$ and include in $B$
all strings $x'$ such that  
either a program $p$ of length 
at most $k$ has halted and output 
a set $C\ni x'$ with $|p|+\log|C|\le k+(2c+1)\log k$,
or we find out that $K(x')<k-2\log k$.
Once $x$ gets in $B$    
we change $x$ to the  first string of length
$k+d\log k$ outside $B\cup S$. 
(We will show that at every step it holds $|B\cup S|< 2^{k+d\log k}$.)

We proceed in this way until  
$A(x,k)$ prints a number $\lambda$ or the number
of changes of $x$ exceed $2^{k+2}$. (Actually, we will prove that
the number of changes of $x$ does not exceed $2^{k+1}+2^{k-2\log k}$.) 
Therefore $K(x)\le k+2\log k$ for all our $x$'s so  
we eventually will find $x$ such that $A(x,k)$ outputs a result
$\lambda$.
If
$\lambda\ge k+(c+1)\log k$ then include $x$ in $S$
and then change $x$  to the  first string of length
$k+d\log k$ outside (the current version of) $B\cup S$. 
Otherwise, when $\lambda<k+(c+1)\log k$, let $\tilde \lambda_x$ be
the current approximation of $\lambda_x$.
We know that $x$ is outside all known sets $C$ with
$K(C)\le k$, $K(C)+\log|C|\le k+(2c+1)\log k$.
Therefore,  for
every $\alpha\le k$ it holds $\tilde \lambda_x(\alpha)>k+(2c+1)\log k$
and hence $|\tilde \lambda_x(\alpha)-\lambda|>c\log k$.
This implies  that either
$K(x)< k-2\log k$ or $\tilde \lambda_x$ differs from $\lambda_x$.
So we are sure that at least one more program of length $k$ or less
still has to halt. We wait 
until this happens, then include $x$ in $B$
and change $x$ to the first string of length $k+d\log k$ outside
$B\cup S$.

Once we get $2^{k-2\log k}$ elements in $S$ we
halt.
Every change
of $x$ is caused by a halting of a new program of length at most $k$
or by including $x$ 
in $S$, thus the total number of changes does not exceed 
$2^{k+1}+2^{k-2\log k}$. 

Note that at every step we have
	$$
|B\cup S|\le 2^{k-2\log k}+2^{k+(2c+1)\log k}+2^{k-2\log k}< 2^{k+d\log k}
	$$
provided that $d>2c+1$. 
	\end{proof}

What if the algorithm is required to approximate $\lambda_x$ only if it
is given the precise value of $K(x)$? We are able to prove that 
in this case the algorithm cannot compute $\lambda_x(\alpha)$ too.
It is even impossible to approximate the
complexity of minimal sufficient statistic.
To formulate this result precisely consider the following
promise problem:

{\bf Input:} $x,k=K(x),\alpha\in[\eps,k-\eps]$.

{\bf Output:} 

1, if $\lambda_x(\alpha-\eps)< k+6\log k$,

0, if $\lambda_x(\alpha+\eps)> k+3\eps$.

If neither of two above cases occurs the algorithm may output
any value or no value at all.

        \begin{theorem}\label{th67}
There is no algorithm $A$ solving this
promise problem for all $x$ and
$\eps=|x|/10\log |x|$.
        \end{theorem}
        \begin{corollary}
There is no algorithm that given
$x,k=K(x)$ finds an integer valued function $\lambda$
on $[0,k]$ such that $\close{\lambda_x}{\lambda}$
for $\eps=\delta=|x|/10\log |x|$.
         \end{corollary}

Indeed, if there were such algorithm we could solve
the above promise problem by answering 1
when $\lambda(\alpha)\le k+2\eps$ and 0 otherwise.

        \begin{proof}
The proof is by contradiction. The idea is as follows. Fix large
$k$. We consider $N=O(\log k)$ points $\alpha_1,\dots,\alpha_N$
that divide the segment $[0,k]$ into equal parts. We
lower semicompute $\lambda_x$ and $K(x)$ for different $x$'s of
length about $k+4\eps$. We are interested in strings $x$ with
$\tilde\lambda_x(\alpha_1+\eps)>k+3\eps$ where $\tilde \lambda_x$
is the current approximation to $\lambda_x$. By counting
arguments there are many such strings. We apply the algorithm to
$\pair{x,\tilde K(x),\alpha_1}$ for those $x$'s, where $\tilde
K(x)$ stands for the currently known upper bound for $K(x)$. Assume
that $A(x,\tilde K(x),\alpha_1)$ halts. If the answer is 1
then we know that $K(x)<\tilde K(x)$ or
$\lambda_x(\alpha_1+\eps)<\tilde\lambda_x(\alpha_1+\eps)$ and we
continue lower semicomputation until we get know which of two
values $\tilde K(x)$ or $\tilde\lambda_x(\alpha_1+\eps)$ gets smaller. If
the latter is decreased we just remove $x$ (the total number
of removed $x$ will not increase $2^{k+3\eps}$ and thus they form a small
fraction of strings of length $k+4\eps$). If for many $x$'s
the answer is 0 we make those answers incorrect by including
those $x$'s in a set of cardinality $2^{k-\alpha_1+2\eps}$ and
complexity $\alpha_1-\eps$. Then for all such $x$'s
$\lambda_x(\alpha_1-\eps)<k+\eps$ and
thus algorithm's answer is incorrect. Hence
$K(x)<\tilde K(x)$ and we continue lower semicomputation. For
all those $x$'s for which $\tilde K(x)$ is decreased we repeat
the trick with $\alpha_2$ in place of $\alpha_1$. In this way we
will force $\tilde K(x)$ to decrease very fast for many $x$'s.
For most of $x$'s $\tilde K(x)$ will become much less than
$k$, which is impossible.

Here is the detailed construction.
Fix large $k$.
Let $N=3\log k$, $\delta=k/9\log k$
(one third of the
distance between consecutive $\alpha_i$),
$\alpha_i=k-3\delta i+\delta$,
$n=k+4\delta$ (the length of $x$).
The value of parameter $\eps$ is chosen to be slightly less
than $\delta$
(we will need that $\delta>\eps+4\log k$ for large enough $k$).

We will run all the programs of length at most $k'=k+2\log k$
and the algorithm $A$ on all possible inputs in a dovetailed fashion.

We will define a set $X$ of $2^k$ strings of length $n$.
Our action will be determined by $k$ only, hence
$K(x)\le k+2\log k=k'$ for every $x\in X$ provided $k$ is large enough.
We will also define some small sets $B_l$  for $l=1,\dots,N$,
the sets of ``bad'' strings and $B$ will denote their union.
Every $B_l$ will have at most $2^{k-\delta}$ elements.
We start with $B_l=\emptyset$ for $l=1,\dots,N$.

We make $2^k$ stages.
At every stage consider the sets
        \begin{align*}
X_l&=\{x\in X\setminus B\colon \tilde K(x)=k'+1-l\}\quad
\text{ for } l=1,\dots,N,\\
X_0&=\{x\in X\setminus B\colon \tilde K(x)>k'\}.
        \end{align*}
Before and after every stage the following invariant will be true.

        \begin{itemize}

\item[(1)] 
$|X_l|<2^{3\delta l}$ for every $0\le l\le N$;
in particular $X_0=\emptyset$.

\item[(2)] For all $1\le l\le N$ for all $x\in X_l$
it holds  $A(x,\tilde K(x),\alpha_l)=0$.

\item[(3)] For all $0\le l<i\le N$ and all $x\in X_l$
it holds $\tilde\lambda_x(\alpha_i+\delta)>k+3\delta$.

\item[(4)] $|B_i| \le 2^{3i\delta-3\delta+1}\times$(the number of
programs of length at most $k-3i\delta+2\delta$ that have halted so far).
        \end{itemize}

At the start all $X_l$'s and $B_i$'s are empty so the invariant
is true. Each stage starts by including a new element in $X$.
This element is the first string $x_0$ of length $n=k+4\delta$ outside $X$
such that $\tilde\lambda_x(\alpha)>k+3\delta$ for all $\alpha\le k$.
    Thus by the choice of $x_0$ the assertions (3) and (4)
    remain true but (1) and (2) may not.

We claim that continuing the dovetailing and updating properly $B_i$'s
we eventually make every one of (1), (2), (3) and (4) true.
During the dovetailing the sets $X_l$  change (an element can move from
$X_l$ to $X_i$ for $i>l$ and even to $X\setminus(X_0\cup\dots\cup X_N)$).
We will denote by $\tilde X_l$ the version of $X_l$
at the beginning of the stage (and $\tilde X_0=\{x_0\}$)
and keep the notations $X_l,B_i$ for current versions
of $X_l,B_i$, respectively.
The rule to update $B_i$'s is very simple:
once at some step of the dovetailing a new set $C$ of complexity
at most $k-3i\delta+2\delta=\alpha_i+\delta$
appears, we include in $B_i$ all the elements of the set
$\bigcup_{j=0}^{i-1}\tilde X_j$.
As
$X_l\subset \bigcup_{j=0}^{l}\tilde X_j$
this keeps (3) true.
Moreover, this keeps true also the following
assertion:
        \begin{itemize}
\item[(5)] For all $1\le l\le N$ for all $x\in X_l\setminus\tilde X_l$
it holds $\tilde\lambda_x(\alpha_l+\delta)>k+3\delta$.
        \end{itemize}
And this also keeps (4) true since
$\bigcup_{j=0}^{i-1}|\tilde X_j|<1+\sum_{j=1}^{i-1}2^{3\delta j}
<2^{3\delta(i-1)+1}$.

We continue the dovetailing and update $B_i$'s as described until
both
$(1)$ and  $(2)$ are true.
Let us prove that this happens eventually.
It suffices to show that if  $(3)$, (4) and (5) are true
but $(2)$ is not, or $(2)$, (3), (4)           and (5)
are true but $(1)$ is not then at least one program of length
$\le k'$ will halt or $A(x,\tilde K(x),\alpha_l)$ is undefined
for some $l$ and some $x\in X_l$.

Consider the second case: $(2)$, (3), (4) and (5)
are true but $(1)$ is not.
Pick $l$ such that $|X_l|\ge 2^{3\delta l}$.
If $l=0$, that is, $\tilde K(x_0)>k'$, we are done, as
$K(x_0)\le k'$.
Otherwise,
let $S$ consist
of the first $2^{3\delta l}$ elements in $X_l$.
We claim that $K(S)\le k-3l\delta+4\log k\le\alpha_l-\eps$.
To prove the claim
we will show that all $S\subset  X_l$
obtained in this way are pairwise disjoint,
therefore their number is at most $2^k/2^{3l\delta}$.
Thus $S$ may be identified by $k,l$ and its index among
all such $S\subset  X_l$.

Therefore                                      for all $x\in S$
we have $\lambda_x(\alpha_l-\eps)<k+4\log k<\tilde K(x)+6\log\tilde K(x)$
and the value
$A(x,\tilde K(x),\alpha_l)=0$ is not
correct. This implies that
$\tilde K(x)$ is not correct for all $x\in S$.
We continue
the dovetailing until all elements of $S$ move
outside $X_{l}$.
Then $S$ becomes disjoint with $X_0\cup\dots\cup X_{l}$
and therefore it will be disjoint with all future
versions of $X_l$.

Consider the first case: $(3)$, (4) and (5) are true
but $(2)$ is not.
Pick $l$ and $x\in X_l$ such that
$A(x,\tilde K(x),\alpha_l)$ is undefined or
$A(x,\tilde K(x),\alpha_l)=1$.
If $A(x,\tilde K(x),\alpha_l)$ is undefined
then we are done: since $\tilde\lambda_x(\alpha_l+\eps)\ge
\tilde\lambda_x(\alpha_l+\delta)>k+3\delta>\tilde K(x)+3\eps$,
either $\tilde\lambda_x$ or $\tilde K$ will decrease,
or $A(x,\tilde K(x),\alpha_l)$ will get defined.
Consider the other case.
Obviously  $x\not\in X_l\setminus\tilde X_l$.
By (5) we have 
$\tilde\lambda_x(\alpha_l+\eps)\ge\tilde\lambda_x(\alpha_l+\delta)>k+3\delta
\ge\tilde K(x)+3\eps$.
Therefore
$\lambda_x(\alpha_l+\delta)<\tilde\lambda_x(\alpha_l+\delta)$ or
$\tilde K(x)<K(x)$ and we are done.

After $2^k$ stages the set
$|X|$ has $2^k$ elements and we have a contradiction.
Indeed, all $X_1,\dots,X_N$ form a very small
part of $X$ because of (1).
The sets $B_1,\dots,B_N$ together form also a very small part
of $X$ because of (4). Thus for most strings $x\in X$ it holds
$\tilde K(x)<k'-N+1\ll k$ which is a contradiction.
        \end{proof}

        \begin{remark}
\rm
Let us replace in the above promise problem
$K(x)$,
the prefix complexity  of $x$, by 
$C(x)$, the plain complexity
of $x$. For the modified problem
we can strengthen the above theorem
by allowing $\eps=|x|/c$ where the constant $c$
depends on the reference computer. Indeed for every $x\in X$ we
have $C(x)\le k+O(1)$: every $x\in X$ can be described by
its index in $X$ in exactly $k$ bits and the value of $k$ may be retrieved
from the length of the description of $x$.
Therefore we will need $N=O(1)$ to obtain a contradiction.
        \end{remark}

After a discussion of these results, Andrei A. Muchnik suggested, and proved,
that if we are also given an $\alpha_0$ such that
$\lambda_x (\alpha_0)\approx K(x)$ but
$\lambda_x (\alpha)$ is much bigger  than $K(x)$ for $\alpha$ much less than
$\alpha_0$
(which is therefore the complexity of the minimal sufficient statistic),
then we can compute
$\lambda_x$ over all of its domain.
This result underlines the significance of the information
contained in the {\em minimal}
sufficient statistic:

	      \begin{theorem}
	      \label{theo.muchnik}
There are a constant $c \geq 0$ and an algorithm that
given any $x,k,\alpha_0$ with
$K(x)\le k\le \lambda_x(\alpha_0)$
finds a non-increasing function $\lambda$ defined on $[0,k]$
such that
$\close{\lambda_x}{\lambda}$ with $\delta = \lambda_x(\alpha_0)-K(x)+O(1)$ 
and
$\eps = \alpha_0-\alpha_1+c\log k$
where $\alpha_1=\min\{\alpha :
\lambda_x(\alpha)\le k+c\log k\}$.
	       \end{theorem}
	       \begin{proof}
The algorithm is a follows.
Let $D_k=\{\pair{y,i}\mid K(y)\le i\le k\}\subset D$.
Enumerate pairs $\pair{y,i}\in D_k$
until a pair $\pair{x,i_0}$ appears
and form a list of all enumerated pairs.
For $\alpha\le\alpha_0$
define $\lambda(\alpha)$ to be the minimum $i+\log|S|$ over
all $S\ni x$ such that a pair $\pair{x,i}$ with $i\le\alpha$
is in the list.
For $\alpha_0< \alpha\le k$
let $\lambda(\alpha)=k$.

For every $\alpha>\alpha_0$ we have
$\lambda_x(\alpha)\ge K(x)-O(1)\ge k-\lambda_x(\alpha_0)+K(x)-O(1)=
\lambda(\alpha)-\delta$
and
$\lambda_x(\alpha)\le \lambda_x(\alpha_0)\le
k+(\lambda_x(\alpha_0)-K(x))\le \lambda(\alpha)+\delta$.

For every $\alpha\le \alpha_0$ we have
$\lambda_x(\alpha)\le\lambda(\alpha)$.
So it remains to show that
for every $\eps\le\alpha\le \alpha_0$
we have $\lambda(\alpha)\le\lambda_x(\alpha-\eps)+\delta$.
We will prove a stronger statement:
$\lambda(\alpha)=\lambda_x(\alpha)$ for every
$\alpha\le\alpha_0-\eps$ provided
$c$ is chosen appropriately.
To prove this it suffices to show
that all for all $S$ with $K(S)\le\alpha_0-\eps$
the pair $\pair{S,K(S)}$ belongs to the list.

By Theorem~\ref{th90} Item (i) we have
$\lambda_x(|m^k_x|+c_1\log k)\le k+c_2\log k$.
That is,
$\alpha_1\le |m^k_x|+c_1\log k$ if
$c\ge c_2$
and
$\alpha_0-\eps=\alpha_0-\alpha_0+\alpha_1-c\log k\le
|m^k_x|+(c_1-c)\log k$.

{}From the proof of Theorem~\ref{th91} we see that there is a constant
$c_3$ such that for every $y$ with
$K(y)\le|m^k_x|-c_3\log k$ the index of 
$\pair{y,K(y)}$ in the enumeration of $D_k$
has less than $|m^k_x|$ common bits with $N^k$.
Assuming that $c\ge c_1+c_3$ we obtain that
the indexes of all pairs $\pair{y,K(y)}$ with
$K(y)\le\alpha_0-\eps$ in the enumeration of $D_k$
are less than $I^k_x$.
		  \end{proof}

\subsection{Randomness Deficiency Function}
The function $\beta_x (\alpha)$ is computable from $x, \alpha$ given an
oracle for the halting problem: run all programs of length $\leq \alpha$
dovetailed fashion and find all finite sets $S$ 
containing $x$ that are produced. 
With respect to all these sets determine the conditional complexity
$K(x \mid S^*)$ and hence the randomness deficiency $\delta (x \mid S)$.
Taking the minimum we find $\beta_x (\alpha)$. All these things are
possible using information from the halting problem to determine
whether a given program will terminate or not.
It is also the case that the function
$\beta_x (\alpha)$ is upper semi-computable from $x,\alpha, K(x)$ up
to a logarithmic error: this follows from the semi-computability of
$\lambda_x (\alpha)$ and Theorem~\ref{th13}.
More subtle is that $\beta_x$ is not semi-computable, not even within
a large margin of error:

\begin{theorem} \label{th.semic}
The function $\beta_x(\alpha)$ is 

(i) not lower semi-computable to within precision $|x|/3$; and

(ii) not upper semi-computable
to within precision $|x|/\log^4|x|$.
\end{theorem}

\begin{proof}
(i) The proof is by contradiction.
Assume Item (i) is false. Choose an arbitrary length $n$.
Let $\beta$ be a function defined by $\beta (i) = \frac{n}{2}$
for $0 \leq i \leq \frac{n}{3}$, and equal 0 otherwise.
Then the function $\beta_x$ with $x$ of length $n$, corresponding to
$\beta$, by Corollary~\ref{cor.beta}, has $x$ with $k=K(x)$ satisfying
$\beta (0) = n-k \pm O(\log n)$ so that $k = \frac{n}{2} \pm O(\log n)$.
Moreover, $\beta_x (i) = \frac{n}{2} \pm O(\log n)$ for
$O(\log n) < i \leq \frac{n}{3}-O(\log n)$, and $\beta_x (i) = O(\log n)$
for $i > \frac{n}{3}+O(\log n)$. Write the set of such $x$'s as $X$.
By dovetailing the lower approximation of $\beta_x (i)$
for all $x$ of length $n$ and some $i$
with $\frac{n}{8} \leq i \leq \frac{n}{4}$,
by assumption on lower semi-computability of $\beta_x$,
we must eventually find an $x$, if not $x \not\in X$ then $x \in X$, 
for which the lower semi-computation of $\beta_x (i)$
exceeds $\frac{n}{2}-\frac{n}{3}- O(\log n)$. But we know
from Corollary~\ref{cor.beta} that $\beta_x (i) = O(\log n)$
for $i > K(x) + O(\log n)$, and hence we have determined
that $i- O(\log n ) < K(x)$. Therefore, $K(x) > \frac{n}{8} - O(\log n)$.
    But this contradicts the well-known fact \cite{LiVi97}
    that there is no algorithm that for any given 
    $n$ finds a string of complexity at least $f(n)$ where
    $f$ is a computable total unbounded function.

(ii) The proof is by contradiction. Assume Item (ii) is false.
Fix a large length
$n=2^k$ and let $A_1= \{0,1\}^n$, so that
$\alpha=2\log k\ge K(A_1)$.
Let $x$  be a string of length $n$, let $N^\alpha<2^{\alpha+1}$
be the number of halting programs of length at most $\alpha$, and
let ${\cal A} =  \{A_1, \ldots , A_m\}$
be the set of all finite sets of complexity at most $\alpha$.
Since $x\in A_1$, the value $\beta_x(\alpha)$ is finite and
$
\beta_x (\alpha) = \min_{A \in {\cal A}} \{ \delta (x \mid A) \}.
$
Assuming $\beta_x$ is upper semi-computable, we can run the following
algorithm:

{\bf Algorithm:}
Given $N^{\alpha}$, $\alpha$, and $x$,

{\bf Step 1:}
Enumerate all finite sets ${\cal A}= \{A_1, \ldots , A_m\}$ of complexity
$K(A_i) \leq \alpha$. Since we
are given $N^\alpha, \alpha$ we can list them exhaustively.

{\bf Step 2:}
Dovetail the following
computations simultaneously:

{\bf Step 2.1:}
Upper semi-compute $\beta_x (\alpha)$, for all $x$ of length $n$.

    {\bf Step 2.2:}
    For all $i=1,\dots,m$,
    lower semi-compute $\delta (x \mid A_i)
    = \log |A_i| - K(x \mid A_i)$.

    We write the approximations at the $t$th step as $\beta^t_x (\alpha)$,
    $\delta^t (x \mid A_i)$, and $K^t (x \mid A_i)$, respectively.
    We continue the computation until step $t$ such
    that
    $$
    \beta_x^t (\alpha) < \min_{A \in {\cal A}} \{ \delta^t (x \mid A) \}+
    n/\log^4n.
    $$
    This $t$ exists by the assumption above.
    By definition, $\min_{A \in {\cal A}} \{\delta(x \mid A)\} =
    \beta_x (\alpha) \leq \beta_x^t (\alpha)$.
    Let
    $A^x$ denote the set
    minimizing the right-hand side.
    (Here we use that
    $x$ belongs to a set in $\cal A$.)
    Together, this shows that $\log |A^x| - \beta_x^t (\alpha)
    \leq K(x \mid A^x)$ and $\log |A^x| - \beta_x^t (\alpha)
    \geq K^t(x \mid A^x) - n/ \log^4 n \geq K(x \mid A^x) - n/ \log^4 n)$.
    Thus we obtained an estimation
    $\log |A^x| - \beta^t_x (\alpha)$
    of $K(x  \mid  A^x)$
    with precision $n/\log^4n$.
    We use that $K(x  \mid  A^x)$ is a good approximation to $K(x)$:
    \begin{align*}
    K(x  \mid A^x)-c_1 & \le K(x)\le K(x  \mid A^x)+|A^x|+c_1
    \\& \le K(x  \mid A^x)+\alpha+c_1,
    \end{align*}
    where $c_1$ is a constant.
    Consequently,
            $$
    K(K(x) \mid x)\le
    K(N^{\alpha}, \alpha, K(x)-\log |A^x| + \beta^t_x (\alpha))+c_2.
            $$
    where the constant $c_2$ is the length of a program to reconstruct
    $\alpha,N^{\alpha}$ and $K(x) - \log |A^x| + \beta^t_x (\alpha)
    \leq \alpha + c_1 + n/\log^4 n$,
    and combining this information with the conditional information $x$,
    to compute $K(x)$.
    Observing
    $K(N^{\alpha}) = \alpha - K(\alpha) + O(1)$ by \cite{GTV01},
    and substituting $\alpha = 2 \log \log n$,
    there is a constant $c_3$ such that
            $$
    K(K(x) \mid x ) \leq 2 \log \log n + \log n - 4 \log \log n +c_3.
            $$
    However, for every $n$, we can choose an $x$ of length $n$ such that
    $K(K(x) \mid x) \geq \log n - \log \log n$ by \cite{Ga74}, which gives the
    required contradiction.
    \end{proof}

{\bf Open question.}
Is there a non-increasing (with respect to $\alpha$)
upper semi-computable function $f_x(\alpha)$ such that,
for all $x$, $\close{\beta_x(\alpha)}{f_x(\alpha)}$ for
$\eps=\delta=O(\log|x|)$ (or for
$\eps=\delta=o(|x|)$)?

\section*{Acknowledgment}
We thank Harry Buhrman, Marcus H\"utter, Andrei A. Muchnik, Osamu Watanabe,
 for helpful discussions,
and Tom Cover, P\'eter  G\'acs, and Leonid A. Levin for historical
details about Kolmogorov's original proposal of the structure
function in the early seventies
and about their own unpublished work on the subject.
After the first author questioned Muchnik whether 
the result in Theorem~\ref{th13} was perhaps known before, 
the latter answered
negatively and supplied
an independent proof of it.
He also suggested and proved Theorem~\ref{theo.muchnik}.
Watanabe suggested the question 
treated 
in Section~\ref{ex.recoding}.

\section*{Biographies}
{\sc Nikolai K. Vereshchagin} is
Professor  (Chair of Mathematical Logic and Theory of Algorithms)
at Moscow State University.
He has worked on algorithms in number theory,
decidable theories,
computational complexity,
Kolmogorov complexity. Together with Alexander Shen he
co-authored ``Computable Functions'' and ``Basic Set Theory'',
MCNMO, Moscow, 1999, in Russian, both of
which have been published by the American
Mathematical Society in English translation.

{\sc Paul M.B. Vit\'anyi} is a Fellow
of the Center for Mathematics and Computer Science (CWI)
in Amsterdam and Professor of Computer Science
at the University of Amsterdam.  He serves on the editorial boards
of Distributed Computing (until 2003), Information Processing Letters,
Theory of Computing Systems, Parallel Processing Letters,
International Journal of Foundations of Computer Science,
Journal of Computer and Systems Sciences (guest editor),
and elsewhere. He has worked on cellular automata,
computational complexity, distributed and parallel computing,
machine learning and prediction, physics of computation,
Kolmogorov complexity, quantum computing. Together with Ming Li
they pioneered applications of Kolmogorov complexity
and co-authored ``An Introduction to Kolmogorov Complexity
and its Applications,'' Springer-Verlag, New York, 1993 (2nd Edition 1997),
parts of which have been translated into Chinese,  Russian and Japanese.

\end{document}